\documentclass[manuscript]{aastex}
\usepackage{amssymb}

\begin{document}
\title{Gas Kinematics and Star Formation in the Filamentary IRDC G34.43+0.24}
\author{Jin-Long Xu\altaffilmark{1,2}, Di Li\altaffilmark{1},  Chuan-Peng Zhang\altaffilmark{1,2}, Xiao-Lan  Liu\altaffilmark{1,2}, Jun-Jie Wang\altaffilmark{1,2}, Chang-Chun Ning\altaffilmark{2,3}, and Bing-Gang Ju\altaffilmark{4,5}}
\affil{$^{1}$  National Astronomical Observatories, Chinese Academy
of Sciences, Beijing 100012, China\\
$^{2}$  NAOC-TU Joint Center
for Astrophysics, Lhasa 850000, China \\
$^{3}$Tibet University, Lhasa,  Tibet, 850000, China\\
$^{4}$Purple Mountain Observatory, Qinghai Station, 817000, Delingha, China\\
$^{5}$Key Laboratory of Radio Astronomy, Chinese Academy of Sciences, China
} \email{xujl@bao.ac.cn}
\begin{abstract}
We performed a multiwavelength study toward infrared dark cloud (IRDC) G34.43+0.24.  New maps of $^{13}$CO $J$=1-0 and C$^{18}$O $J$=1-0 were obtained from the Purple Mountain Observatory (PMO) 13.7 m radio telescope.  At 8 $\mu$m (Spitzer - IRAC), IRDC G34.43+0.24  appears to be a dark filament extended by 18$^{\prime}$ along the north-south direction. Based on the association with the 870 $\mu$m and C$^{18}$O $J$=1-0 emission, we suggest that IRDC G34.43+0.24 should  not be 18$^{\prime}$ in length, but extend by 34$^{\prime}$. IRDC G34.43+0.24 contains some massive protostars, UC H II regions, and infrared bubbles. The spatial extend of IRDC G34.43+0.24 is about 37 pc assuming a distance of 3.7 kpc. IRDC G34.43+0.24 has a linear mass density of $\sim$ 1.6$\times$10$^{3}$ $M_{\odot}$ $ \rm pc^{-1}$, which is roughly consistent with its critical mass to length ratio. The turbulent motion may help stabilizing the filament against the radial collapse.  Both infrared bubbles N61 and N62 show a ringlike structure at 8 $\mu$m. Particularly, N61 has a double-shell structure, which has expanded into IRDC G34.43+0.24. The outer shell is traced by 8 $\mu$m and $^{13}$CO $J$=1-0 emission, while the inner shell is  traced by 24 $\mu$m and 20 cm emission. We suggest that the outer shell (9.9$\times10^{5}$ yr) is created by the expansion of H II region G34.172+0.175, while the inner shell (4.1$\sim$6.3$\times$10$^{5}$ yr) may be produced by the energetic stellar wind of its central massive star. From GLIMPSE I catalog, we selected some Class I sources with an age of $\sim$$10^{5}$ yr. These Class I sources are clustered along the filamentary molecular cloud.
\end{abstract}

\keywords{
ISM: clouds---ISM: individual objeccts (G34.43+0.24, N61 and N62)---
Stars: formation ---ISM: H {\footnotesize II} regions}
\section{INTRODUCTION}
Because infrared dark clouds (IRDCs) are  considered the precursors of massive stars and star clusters  (Egan et al. 1998; Carey et al. 2000; Rathborne et al. 2006). They have attracted much attention particularly in the last decade. IRDCs are seen as dark absorption features against the Galactic background at mid-infrared  wavelengths (Egan et al. 1998; Hennebelle et al. 2001; Simon et al. 2006; Peretto \& Fuller 2009). Egan et al. (1998) initially suggested that IRDCs are isolated objects, possibly left over after their parental molecular clouds have been dispersed. Subsequent studies of IRDCs suggest that they are cold ($<$25 K), dense ($\thicksim$10$^{3}$-10$^{5}$ cm$^{-3}$), and are thought to host very early stages of massive star formation (Carey et al. 2000; Simon et al. 2006; Pillai et al. 2006; Ragan et al. 2011; Liu et al. 2014).  One striking feature of IRDCs is their filamentary shape. Jackson et al. (2010) mentioned that such filaments probably resulted from the passage of a spiral shock.  Wang et al. (2015) found that some filaments are closely associated with the Milky Way spiral substructures. In addition, compact cores are found in IRDCs (Rathborne et al. 2006). Some are found to be candidates for massive starless objects (Tackenberg et al. 2012; Beuther et al. 2013), other  show the signposts of massive star formation (Rathborne et al. 2011), such as hot  molecular cores, ultra-compact H {\footnotesize II} (UC H {\footnotesize II}) regions, and infall and outflow (Rathborne et al. 2007, 2008). Another important feature of massive star formation is when a massive star formed inside an IRDC, UV radiation and stellar winds would ionize the surrounding gas and create an infrared bubble (Churchwell et al. 2006), and even disrupt the natal IRDC (Jackson et al. 2010).

Infrared bubbles are the bright 8.0 $\mu$m emission surrounding O and early-B stars, which show the full or partial ring structures (Churchwell et al. 2006). Churchwell et al. (2006 \& 2007) have compiled a list of $\sim$600 infrared bubbles, while Simpson et al. (2012) have created a new catalogue of 5106 infrared bubbles through visual inspection via the online citizen science website  `The Milky Way Project (MWP)'. Deharveng et al. (2010) studied 102 bubbles. They show that 86$\%$ of these bubbles enclose H {\footnotesize II} regions ionized by O and B stars. Since the bubbles are usually found in or near massive star-forming regions, which can provide important information on the dynamical processes and physical environments of their surrounding cloud.

G34.43+0.24 is a filamentary IRDC at a kinematic distance of 3.7 kpc (Rathborne et al. 2006; Fa\'{u}ndez et al. 2004;
Simon et al. 2006). VLBI parallax measurements of H$_{2}$O maser sources within the IRDC determined a distance of 1.56 kpc (Kurayama et al. 2011). Although a parallax measurement would seem to be the more reliable distance determination, Foster et al. (2012)  suggested that the parallax determinations to the same sources are incorrect. Foster et al. (2014) considered that the Kurayama et al. (2011) measurement relied on only a single reference background source, and the measurements are inherently difficult due to the low declination of this target. They thus adopted the kinematic/extinction distance of 3.9 kpc for this cloud. IRDC G34.43+0.24 spans by 9$^{\prime}$ from north to south in equatorial projection to a total mass of 1000 M$_{\odot}$ from the ammonia observation (Miralles et al. 1994). Famous UC H {\footnotesize II} complex G34.26+0.15 is roughly 11$^{\prime}$ south of IRDC G34.43+0.24 (Shepherd et al. 2007).  Rathborne et al. (2006) identified nine cores in this IRDC (see Fig. 1). The MM1 core appears to be a massive B2 protostar in an early stage of evolution, based on the weak 6 cm continuum emission and the lack of detection at NIR wavelengths (Shepherd et al. 2004). Rathborne et al. (2011) confirmed that the MM1 core is a hot molecular core with a rotating structure surrounding the central protostar (Rathborne et al. 2008). From  CO observations with Owens Valley Radio Observatory (OVRO) array of six 10.4 m antenna, two massive outflows were discovered in the MM1 core (Shepherd et al. 2007). The MM2 core is associated with UC H {\footnotesize II} G34.4+0.23 (Miralles et al. 1994; Molinari et al. 1998), IRAS 18507+0121 source (Bronfman et al. 1996), an H$_{2}$O maser (Miralles et al. 1994), and an CH$_{3}$OH maser (Szymczak \& Kus 2000). Shepherd et al. (2007) also detected three massive outflows centered on or near UC H {\footnotesize II} G34.4+0.23. The MM2 core is also undergoing infalling motion with a mass infall rate of about 1.8$\times$10$^{-3}$ M$_{\odot}$ yr$^{-1}$. The MM3 core is located close to  the northern edge of the filament IRDC G34.43+0.24, about 3.5$^{\prime}$ north of the MM2.  Sanhueza et al. (2006) reported an outflow in CO $J$=3-2 toward the MM3 core. From ALMA observations, the 1.3 mm continuum and several molecular lines reveal a highly collimated outflow with a dynamic timescale of less than 740 yr and a hot molecular core toward the MM3 core (Sakai et al. 2013). Although star formation has already started in the MM3 core (Rathborne et al. 2005), the MM3 core is thought to be in an earlier evolutionary stage than the MM1 and MM2 cores. The rest of the cores (MM4, MM5, MM6, MM7, MM8, and MM9) do not have obvious sign of massive star formation.

To investigate the gas kinematics and star formation in IRDC G34.43+0.24, we performed a multi-wavelength study toward the IRDC. Combining  the NRAO VLA Sky survey, ATLASGAL survey, and GLIMPSE survey, we aim to construct a comprehensive
larger-scale picture of IRDC G34.43+0.24.   The observations and data reduction are described in Sect.2, and the results are presented in Sect.3. In Sect.4, we will discuss the star formation scenrio. The conclusions are summarized in Sect.5.

\section{OBSERVATIONS AND DATA REDUCTION}
\subsection{Purple Mountain Data}
We made the mapping observations of IRDC G34.43+0.24 and its adjacent regions in the transitions of $^{13}$CO $J$=1-0 and C$^{18}$O $J$=1-0 lines using the Purple Mountain Observation (PMO) 13.7 m radio telescope at De Ling Ha in the west of China at an altitude of 3200 meters, in 2012 May and 2013 Jan. The new 3$\times$3 beam array receiver system in single-sideband (SSB) mode was used as front end. The back end is a Fast Fourier Transform Spectrometer (FFTS) of 16384 channels with a bandwidth of 1 GHz, corresponding to a velocity resolution of 0.17 km s$^{-1}$
for  $^{13}$CO $J$=1-0 and C$^{18}$O $J$=1-0. The half-power beam width (HPBW)
was 53$^{\prime\prime}$ at 115 GHz and the main beam efficiency was 0.5.  The pointing accuracy of the telescope was better than 5$^{\prime\prime}$,  which was derived from continuum observations of planets. The source W51D (19.2 K) was observed once per hour as flux calibrator. The system noise temperature (Tsys) in SSB mode varied between 150 K and 400. The mean rms noise level of the calibrated brightness temperature was 0.3 K for  $^{13}$CO $J$=1-0 and C$^{18}$O $J$=1-0. Mapping observations were centered at RA(J2000)=$18^{\rm h}53^{\rm m}16.52^{\rm s}$, DEC(J2000)=$01^{\circ}12'43.3^{\prime\prime}$ using the on-the-fly mode with a constant integration time of 6 second at each point. The standard chopper wheel calibration technique is used to measure antenna temperature $T_{\rm A}$$^{\ast}$ corrected for atmospheric absorption. The final data was recorded in brightness temperature scale of $T_{\rm mb}$ (K). The data were reduced using the GILDAS/CLASS \footnote{http://www.iram.fr/IRAMFR/GILDAS/} package.

\subsection{Archival Data }
The 1.4 GHz radio continuum emission data were obtained from the
NRAO VLA Sky Survey  (NVSS; Condon et al. 1998) which is a 1.4 GHz continuum survey
covering the entire sky north of -40$^{\circ}$ declination with a noise of about 0.45 mJy/beam and a resolution of 45$^{\prime\prime}$.

We extract 870 $\mu$m data from the ATLASGAL survey, which is one of the first systematic surveys of the inner Galactic
plane in the submillimetre band. The survey was carried out
with the Large APEX Bolometer Camera (LABOCA; Siringo et al.
2009), an array of 295 bolometers observing at 870 $\mu$m (345 GHz).
The APEX telescope has a full width at half-maximum (FWHM)
beam size of  19$^{\prime\prime}$ at this frequency and has a positional accuracy
of 2$^{\prime\prime}$. The initial survey covered a Galactic longitude
region of 300$^{\circ}$$<$$\ell$$<$60$^{\circ}$ and $|b|$$<$1.5$^{\circ}$ but this was later extended
to include 280$^{\circ}$$<$$\ell$$<$300$^{\circ}$ and -2$^{\circ}$$<$$b$$<$1$^{\circ}$ to take account of
the warp present in this part of the Galactic plane (Schuller et al.
2009).

We also utilize NIR data in four bands from the Spitzer GLIMPSE survey and MIR data in two bands.
GLIMPSE survey (Benjamin et al. 2003) observed the Galactic plane (65$^{\circ}$
$<$ $|l|$ $<$ 10$^{\circ}$ for $|b|$ $<$ 1$^{\circ}$) with the four
IR bands (3.6, 4.5, 5.8, and 8.0 $\mu$m) of the Infrared Array
Camera (IRAC) (Fazio et al. 2004) on the Spitzer Space
Telescope. The resolution ranges from 1.5$^{\prime\prime}$ (3.6 $\mu$m) to
1.9$^{\prime\prime}$ (8.0 $\mu$m).  MIPSGAL is a survey of the same region as GLIMPSE, using MIPS instrument (24 and 70 $\mu$m) on Spitzer (Rieke et al. 2004). The MIPSGAL resolution at 24 $\mu$m is 6$^{\prime\prime}$.

\section{RESULTS}
\subsection{Infrared and Radio Continuum Images of IRDC G34.43+0.24}
Figure 1a shows the Spitzer-IRAC 8 $\mu$m emission of IRDC G34.43+0.24. At 8 $\mu$m, IRDC G34.43+0.24 displays a dark extinction feature from north to south.  Adjacent to  the south of IRDC G34.43+0.24,  we find some bright 8 $\mu$m emission and smaller dark filaments, which is marked in the white dashed lines and seems to be smaller IRDCs. The bright Spitzer-IRAC 8 $\mu$m emission is attributed to polycyclic aromatic hydrocarbons (PAHs) (Leger \& Puget 1984). The PAHs molecules can be destroyed inside the ionized gas, but are excited in the photodissociation region (PDR) by the UV radiation within H {\footnotesize II} region (Pomar\`{e}s et al. 2009). Hence, the Spitzer-IRAC 8 $\mu$m emission can be used to delineate an infrared bubble. Infrared bubbles N61 and N62 are firstly identified by Churchwell et al. (2006). N61 is located to the north of the H {\footnotesize II} complex G34.26+0.15, while N62 is to the south. Both N61 and N62 show a ringlike structure opened at the south.  From the Simpson et al. (2012) catalogue, we  identified another six infrared bubbles in the observed region, named at B1-B6. We use the green circles to represent bubbles B1-B6 in Fig. 1a. The parameters of these bubbles are listed in Table 1. From the position and radius of each bubble, we note that B1-B4 overlap with each other and are situated on the bright 8 $\mu$m clump, while B5 and B6 are located on  the rim of N61 in Fig. 1a.

Figure 1b presents the Spitzer-MIPSGAL 24 $\mu$m image of IRDC G34.43+0.24. IRDC G34.43+0.24 mostly appears as dark extinction at 24 $\mu$m. Rathborne et al. (2006) identified nine dust cores (MM1, MM2, MM3, MM4, MM5, MM6, MM7, MM8, and MM9) in this IRDC (see Fig. 1).  MM1-MM3 are associated with the bright 24 $\mu$m point sources, which are generally considered protostars (Chambers et al. 2009). B1-B6  exhibit the strong 24 $\mu$m emission. Because B1-B4 overlap  with each other in the line of sight, we  cannot distinguish the morphology of the hot dust of each bubble  from each other. In Fig. 1b,  both N61 and N62  contain  24 $\mu$m emission in their inner regions. Moreover, the 24 $\mu$m emission also delineate the ringlike shells of N61 and N62, which is similar to those in the 8 $\mu$m emission. The 24 $\mu$m emission inside N61 shows a semi-ringlike shape, as also shown in the right panel of Fig. 10.

Figure 1c shows the ATLASGAL 870 $\mu$m emission of IRDC G34.43+0.24. The 870 $\mu$m emission traces the distribution of cold dust (Beuther et al. 2012). The long dark filament of IRDC G34.43+0.24 and other smaller IRDC fragments coincide well with the 870 $\mu$m emission.  MM1-MM4 cores, UC H {\small II} region G34.24+0.13, and H {\small II} complex G34.26+0.15 are associated with  bright 870 $\mu$m clumps. UC H {\small II} region G34.24+0.13  hosts a massive protostellar object that coincides with a methanol maser (Hunter et al. 1998).  Taking into consideration the morphology in both 8 $\mu$m and 870 $\mu$m, we put forth the hypothesis that IRDC G34.43+0.24 and the few smaller IRDCs south of IRDC G34.43+0.24 are one continuous structure lying behind the H {\footnotesize II} region G34.172+0.175, with a spatial extent of $\sim$28$^{\prime}$, much longer than the 9$^{\prime}$ listed in Miralles et al. (1994).

The 21 cm emission is mainly from free-free emission, which can be used to trace the ionized gas of H {\small II} regions. Figure 1d presents the 21 cm continuum emission map. The infrared bubbles B1-B6 are associated with  ionized gas, while there is no 21 cm emission from IRDC G34.43+0.24.  In a GBT X-band survey (Anderson et al. 2011), 448 previously unknown Galactic H {\small II} regions were detected  in the Galactic zone 343$^{\circ}$$\leq$$\ell$$\leq$67$^{\circ}$ and $|b|$$\leq$1$^{\circ}$. H {\footnotesize II} regions G34.172+0.175 and G34.325+0.211 are located in our investigative region from the catalog of Anderson et al. (2011).  The hydrogen radio recombination line (RRL) velocities of the two H {\footnotesize II} regions  are 57.3 $\pm$ 0.1 km s$^{-1}$ and 62.9 $\pm$ 0.1 km s$^{-1}$, which are associated with N61 and N62, respectively (Anderson et al. 2011). In Fig. 1d, the ionized gas of G34.325+0.211 shows a compact structure, while it displays a ringlike shape for G34.172+0.175, as shown in a blue dashed circle.

\subsection{CO Molecular Emission of IRDC G34.43+0.24}
IRDCs are dense with the high-volume densities of $\thicksim$10$^{3}$-10$^{5}$ cm$^{-3}$ (Carey et al. 2000; Simon et al. 2006; Ragan et al. 2011). Comparing the median optically thick $^{13}$CO line, the optically thin and relatively abundant C$^{18}$O line (Yonekura et al (2005), whose emission is more suited to trace IRDCs. Figure 2 shows the channel maps of C$^{18}$O $J$=1-0 in step of 1 km s$^{-1}$. The channel maps are superimposed on the Spitzer-IRAC 8 $\mu$m emission. C$^{18}$O is detected  in a velocity range between 53 and 65 km s$^{-1}$ and are well correlated with IRDC G34.43+0.24. Figure 3 displays the integrated intensity map of C$^{18}$O $J$=1-0 overlaid on the Spitzer-IRAC 8 $\mu$m emission. From the perspective of large scale, the C$^{18}$O $J$=1-0 emission shows a filamentary structure extended by $\sim$28$^{\prime}$ from north to south.
Figure 4 presents the velocity-field (Moment 1) map  of C$^{18}$O $J$=1-0 of IRDC G34.43+0.24 overlaid with its integrated intensity contours. From the velocity-field map, we can discern that the filament is in the velocity interval of 53--63 km s$^{-1}$, indicating that the filament is a single coherent object.  To investigate spatial correlation between gas and dust of the filament, we made the $^{13}$CO $J$=1-0 and C$^{18}$O $J$=1-0 integrated intensity maps overlaid on the ATLASGAL 870 $\mu$m emission (Figure 5). The $^{13}$CO $J$=1-0 emission is more extended than that of the ATLASGAL 870  $\mu$m and C$^{18}$O $J$=1-0, whereas the spatial distribution of the C$^{18}$O $J$=1-0 emission is similar to that of the 870 $\mu$m dust emission.

In addition, the Spitzer-IRAC 8 $\mu$m emission can be divided into three regions, named as regions I, II, and III (see Fig. 3). We do not have C$^{18}$O data for the 6$^{\prime}$ north-south  dark filament in region I. Region II contains IRDC G34.43+0.24, which coincides well with the C$^{18}$O $J$=1-0 emission ($\sim$12$^{\prime}$ length). In region III,  a C$^{18}$O $J$=1-0 molecular clump has a dimension of about 16$^{\prime}$ and is correlated with the bright 8 $\mu$m emission and some small IRDCs. The molecular clump contains UC H {\small II} region G34.24+0.13 and H {\small II} complex G34.26+0.15. Moreover, the molecular clump also shows two elongated structures along its northwest. Liu et al. (2013) also identified these elongated structure in the Spitzer-IRAC [4.5]/[3.6] flux ratio map. As depicted by the long white dashed lines, we find that the two elongated structures are coincident with two small IRDCs in the northwest. To the southwestern edge of the molecular clump, the C$^{18}$O $J$=1-0 emission shows a half molecular shell with an opening toward the south. The half shell is associated well with the infrared bubble N61.

To estimate the mass of the coherent filament, we use the optical thin C$^{18}$O emission. Based on the mapped region of the filament in C$^{18}$O $J$=1-0, we only calculate the gas mass from region II to region III. Assuming  the local thermodynamical equilibrium (LTE), the column density are determined by (Scoville et al. 1986)
\begin{equation}
\mathit{N_{\rm C^{18}O}}=4.75\times10^{13}\frac{T_{\rm ex}+0.88}{exp(-5.27/T_{\rm ex})}\int T_{\rm mb}dv ~\rm cm^{-2},
\end{equation}
where $dv$ is the velocity range in km s$^{-1}$, $T_{\rm mb}$ is the corrected main-beam temperature of C$^{18}$O
$J$=1-0, and $T_{\rm ex}$ is the excitation temperature of molecular gas. Region II is a IRDC, we adopt an excitation temperature of 10 K in this region (Simon et al. 2006). For the region III containing H {\small II} regions, an excitation temperature of 20 K  is adopted (Dirienzo et al. 2012).  The C$^{18}$O abundance of $N(\rm H_{2})/N(C^{18}O)$ is about 7$\times10^{6}$ (Castets et al. 1982). The mean number density of $\rm H_{2}$ is estimated to be
\begin{equation}
\mathit{n(\rm H_{2})}=8.1\times10^{-20}N(\rm H_{2})/r,
\end{equation}
where $\rm r$ is the averaged radiuses of regions II and III in parsecs (pc). Their mass is given by
\begin{equation}
\mathit{M_{\rm H_{2}}}=\frac{4}{3}\pi
r^{3}\mu_{g}m(\rm H_{2})n(\rm H_{2}),
\end{equation}
where $\mu_{g}$=1.36 is the mean atomic weight of the gas, and $m(\rm
H_{2})$ is the mass of a hydrogen molecule. The obtained parameters of region II and region III are listed in Table 1. The total mass of gas  is $\sim$4.8$\times10^{4}$$\rm M_{ \odot}$ from region II to region III.

\subsection{Molecular Clump with H {\small II} Complex G34.26+0.15}
The molecular clump in region III contains H {\small II} regions and infrared bubbles. We analyze the gas dynamics in the H {\small II} regions and bubbles using $^{13}$CO $J$=1-0. Figure 6 shows the channel maps of $^{13}$CO $J$=1-0 in the velocity range of 48--67 km s$^{-1}$ with interval 1 km s$^{-1}$. By comparison with the Spitzer-IRAC 8 $\mu$m emission, we identify two half shells of $^{13}$CO $J$=1-0 gas at 48--53 km s$^{-1}$ and 56--60 km s$^{-1}$. There are also two dense cores that can be identified by the channel maps. The peak positions of the identified cores are marked by  the white crosses in Fig. 6. The velocity range of one core is from 52  km s$^{-1}$ to 56  km s$^{-1}$, the other is in the 53--66 km s$^{-1}$ interval. Adopting above four velocity ranges, we made the integrated $^{13}$CO $J$=1-0 maps overlaid on the Spitzer-IRAC 8 $\mu$m image, shown in  Figures 7(a), 7(b), 7(c) and 7(d), respectively. In  Fig. (7a), comparing the scales of infrared bubbles B1, B2, B3, and B4 with that of the half shell,  B1 is  consistent with the half shell of the molecular gas at 48--53 km s$^{-1}$. Because the $^{13}$CO $J$=1-0 component of the central velocity at 51 km s$^{-1}$ also show clearly a half shell (green contours) in  Fig. 6,  we adopt 51 km s$^{-1}$ as the velocity of  B1. In  Fig. 7(b), the northwest core is associated with H {\small II} complex G34.26+0.15. We did not find the associated objects with the southeast core, but find that B2 is located between two cores. Hence, B2 may be coincident with the molecular gas at 52--56 km s$^{-1}$. In  Fig. 7(c), the ATLASGAL 870 $\mu$m emission of G34.26+0.15 shows a compact core, which is well coincident with the northwest core identified in  Fig. 7(b). Previous centimeter observations showed that the complex G34.26+0.15 consists of two hypercompact H {\small II} regions called at A and B, a cometary ultracompact H {\small II} region named C, and an extended ringlike H {\small II} region called component D (Reid \& Ho 1985). In  Fig. 7(d), the $^{13}$CO $J$=1-0 half shell is associated well with N61. The molecular gas shows two integrated intensity gradients toward G34.172+0.175 and infrared bubble B3, as marked by the white arrows. Anderson et al. (2011) found that the RRL velocity of G34.172+0.175 is 57.3 $\pm$ 0.1 km s$^{-1}$, which is just located between 56 and 60 km s$^{-1}$. Hence, G34.172+0.175 is interacting with the half molecular shell. Figure 6 shows that  N62 is consistent with the molecular gas of 56 to 60 km s$^{-1}$ in the morphology.

Figure  8 displays the spectra of $^{13}$CO $J$=1-0 and C$^{18}$O $J$=1-0 toward H {\small II} complex G34.26+0.15. The $^{13}$CO $J$=1-0 line  shows asymmetric profile with double peaks, which may be caused by large optical depths (Allen et al.  2004). The C$^{18}$O $J$=1-0 line with a single emission peak can be considered as optically thin, hence it can be used to determine the systemic velocity. We measure a systemic velocity of $\sim$ 58.2 $\pm$ 0.1 km $\rm s^{-1}$ for G34.26+0.15, which is associated with the hydrogen RRL velocity (57.3 $\pm$ 0.1 km s$^{-1}$) of G34.172+0.175 (Anderson et al. 2011). The $^{13}$CO $J$=1-0 spectrum shows a blue-profile signature (Wu et al. 2007), and the blue peak of $^{13}$CO $J$=1-0 is stronger than its red peak with an absorption dip near the systemic velocity, which may be produced by infall motion or accretion.
In order to identify which one creates the spectral signature, we plot the map grids of H {\small II} complex G34.26+0.15 using $^{13}$CO $J$=1-0, presented in Fig.  9. The map grids indeed exhibit the large-scaled blue asymmetric feature. Moreover, Liu et al. (2013) detected an infall motion with the HCN $J$=3-2, HCO$^{+}$ $J$=1-0, and  CN $J$=2-1 lines for G34.26+0.15 from the JCMT and SMA observations. They obtained the mass infall rate of 1.2$\times$10$^{-2}$ M$_{\odot}$ yr$^{-1}$. Hence, the infall motion in G34.26+0.15 may be responsible for signature of the detected $^{13}$CO $J$=1-0 spectrum.

\subsection{Infrared Bubbles N61 and N62}
Infrared bubbles N61 and N62 lie on the borders of a filamentary molecular cloud in the line of sight.  To clearly analyze the morphology of N61 and N62, we present the zoomed three color images from the Spitzer 3.6 $\mu$m, 8 $\mu$m, and 24 $\mu$m bands for the two bubbles, as shown in Fig.  10.  The white contours correspond to the 21 cm continuum emission extracted from NVSS. In Fig. 10, both the left and right panels  show the PDR visible in the 8 $\mu$m emission, which originates mainly in the PAHs. The PAHs emission of both N61 and N62 displays an almost ringlike shape with an opening towards the south. The openings of the two bubbles resemble a typical `champagne flow' (Stahler \& Palla 2005) with lower density material at the south. Both the 24 $\mu$m and 21 cm continuum emission are visible within the two bubbles. The 21 cm continuum emission traces the ionized gas from H {\footnotesize II} regions G34.172+0.175 and G34.325+0.211 (Anderson et al. 2011). Interestingly, the 24 $\mu$m and 21 cm continuum emission in N61 presents a central cavity. Deharveng et al. (2010) studied 102 bubbles. Ninety-eight percent of these bubbles exhibit 24 $\mu$m emission in their central regions. Only bubble N49 show a central hole in both the radio continuum emission and  the 24 $\mu$m emission (Watson et al. 2008).  They suggested that a central hole could be attributed to a stellar wind emitted by the central massive star. The CO molecular shell associated with N61 exhibits an integrated intensity gradient toward the filamentary molecular cloud. We suggest that H {\footnotesize II} region G34.172+0.175 has expanded into the molecular gas, and the produced shocks have collected the molecular gas into a molecular shell.

To estimate the  age of the molecular shell in the filamentary molecular cloud, we calculate the dynamical age of H {\small II} region G34.172+0.175. Because H {\small II} region G34.325+0.211 that created N62 is close to the filamentary molecular cloud, we also  computed the dynamical age of this H {\small II} region. Assuming an H {\small II} region expanding in a homogeneous medium,  the dynamical age is estimated by (Dyson \& Williams 1980)
\begin{equation}
\mathit{t_{\rm  H {\small II}}}=7.2\times10^{4}(\frac{R_{\rm H {\small II}}}{\rm pc})^{4/3}(\frac{Q_{\rm Ly}}{10^{49} \rm ph~s^{-1}})^{-1/4}(\frac{n_{\rm i}}{10^{3}\rm cm^{-3}})^{-1/2} \rm ~yr,
\end{equation}
where $R_{\rm H {\small II}}$ is the radius of H {\small II} region, $n_{\rm i}$ is the initial number density of gas, and $Q_{\rm Ly}$ is the ionizing luminosity.

Assuming the radio continuum emission is optically thin, the ionizing luminosity $Q_{\rm Ly}$ is computed by Condon (1992)
\begin{equation}
 \mathit{Q_{\rm Ly}}=7.54\times10^{46}(\frac{\nu}{\rm GHz})^{0.1}(\frac{T_{e}}{\rm K})^{-0.45}(\frac{S_{\nu}}{\rm Jy})(\frac{D}{\rm kpc})^{2}\rm ~s^{-1},
\end{equation}
Where $\nu$ is the frequency of the observation, $S_{\nu}$ is the observed specific flux density, and $D$ is the distance to the H {\small II} region. H {\small II} regions G34.172+0.175 and G34.325+0.211 have the flux density of 1.7$\pm$0.7 Jy and (3.1$\pm$0.4)$\times$10$^{-1}$ Jy at 9 GHz (Anderson et al. 2011), respectively. We adopt an effective electron temperature of 10$^{4}$ K, and a distance of 3.7 kpc (see Sect.4.1). Finally,  we obtain $Q_{\rm Ly}$$\backsimeq$(3.5$\pm$1.5)$\times10^{46}$ ph s$^{-1}$ and (6.3$\pm$0.8)$\times10^{45}$ ph s$^{-1}$ for G34.172+0.175 and G34.325+0.211.
Inoue et al. (2001) suggested that only half of Lyman continuum photons from  the central source in a Galactic  H {\small II} region ionizes neutral hydrogen, remainder being absorbed by dust grains within the ionized region.  Using Smith et al. (2002), we determined the spectral type of the ionizing star of G34.172+0.175 is between B0.5V and B1V, while between B1V and B1.5V for G34.325+0.211.

G34.172+0.175 and G34.325+0.211 are located on the borders of a filamentary cloud. We assume that the ionized stars of the two H {\small II} regions may form in a filamentary IRDC.  Simon et al. (2006) has mapped 379 IRDCs in the $^{13}$CO $J$=1-0 molecular line with the Boston University-Five College Radio Observatory Galactic Ring Survey. They obtained a volume-averaged H$_{2}$ densities of $\sim$2$\times$10$^{3}$ cm$^{-3}$ for all the IRDCs. We take the volume-averaged H$_{2}$ as the initial number density of the gas. Adopting the radius of $\sim$3.2$\pm$0.5 pc and $\sim$1.5$\pm$0.1 pc for G34.172+0.175 and G34.325+0.211 obtained from Anderson et al. (2011), we derived that the  dynamical ages of H {\small II} regions G34.172+0.175 and G34.254+0.14 are (9.9$\pm$0.2)$\times10^{5}$ yr and 5.6$\pm$0.1)$\times10^{5}$ yr, respectively. Because H {\small II} regions spend a significant amount of time in the ultra-compact and compact phases, our obtained ages for the H {\small II} regions may be lower limits.

In H {\small II} region G34.172+0.175,  there is a cavity of ionized gas created by the stellar wind from its central massive stars. Using the stellar type of the central massive star, we can estimate the timescale of the ionized cavity.  Applying the equation (McCray et al. 1983)
\begin{equation}
\mathit{R_{W}}=4.3\times10^{-10}(L_{W}/n_{0})^{1/5}t^{3/5}_{W},
\end{equation}
where $n_{0}~[\rm cm^{-3}]$ is  the gas medium density, $L_{W}~[\rm ergs/s]$ is the wind luminosity, and $R_{W}~[\rm pc]$ is the cavity radius. The wind luminosity can be expressed as (Castor et al. 1975)
\begin{equation}
\mathit{L_{W}}=3.16\times10^{-35} \dot{M}_{W}v_{W}^{2},
\end{equation}
where $\dot{M}_{W}$ and $v_{W}$ are the mass-loss rate and velocity of the stellar wind, respectively. The spectral type of the central star is B1V$\sim$B0.5V for G34.172+0.175. Adopting a mass-loss rate and velocity of the stellar wind of B0.5V$\sim$B1V star summarized by Chen et al. (2013), we obtain the wind luminosity ($L_{W}$) of 0.9$\sim$3.3$\times$10$^{33}$ erg/s.  We measured that the cavity radius is about 1.2 pc (1.1$^{\prime}$ at 3.7 kpc) from  Fig. 10 (left panel).
Using the medium density of $\sim$2$\times$10$^{3}$ cm$^{-3}$, we infer that the timescale of the ionized cavity is 4.1$\sim$6.3$\times$10$^{5}$ yr, which is much less than the main-sequence (MS) lifetime of such stars (1.3$\sim$1.6$\times$10$^{7}$ yr for B1V$\sim$B0.5V stars, Chen et al. 2013).
Comparing the timescale of the ionized cavity with that of the infrared bubble N61 (9.9$\pm$0.2$\times10^{5}$ yr), we suggest that  the ionized cavity has been blown via the energetic stellar wind after H {\footnotesize II} region G34.172+0.175 begin to expand in a filamentary molecular cloud. Similarly, after some time an ionized cavity will emerge in infrared bubble N62.

\subsection{Distribution of Young Stellar Objects}
A population of low-mass stars with an age of about a few Myr were detected in IRDC G34.43+0.24 (Shepherd et al. 2007).  In this section, we examine the young stellar object (YSO) populations by GLIMPSE I catalog toward  IRDC G34.43+0.24.  From the catalog, we selected 11872 near-infrared sources with the 3.6, 4.5, 5.8, and
8.0 $\mu$m, within a circle of 23$^{\prime}$ in radius centered on
R.A.=18$^{\rm h}53^{\rm m}15.054^{\rm s}$ (J2000),
Dec=$+01^{\circ}19'19.81^{\prime\prime}$ (J2000). The size of this
region completely covers the extension of IRDC G34.43+0.24 and its south region.   Figure 11 shows the $[5.8]-[8.0]$ versus $[3.6]-[4.5]$ color-color (CC) diagram. The regions in the figure indicate the stellar evolutionary stages based on the criteria of
Allen et al. (2004),  Paron et al (2009), and
Petriella et al (2010). YSOs are generally classified according to their evolutionary stage. These near-infrared sources
are classified into three regions: Class I sources are protostars
with circumstellar envelopes, Class II sources are
disk-dominated objects, and other sources. Class I sources
occur in a period on the order of $\sim$$10^{5}$ yr, while the age
of Class II sources is $\sim$$10^{6}$ yr (Andr\'{e} \& Montmerle
1994).  Using this criteria, we find 256 Class I sources and 419 Class II sources. Here
Class I and Class II sources are chosen to be YSOs.

Figure 12 (left panel) shows the spatial distribution of both Class I and Class
II sources.  From Fig. 12 (left panel), we note that Class I sources (red dots) are asymmetrically distributed across the whole selected region, and are mostly concentrated in the large-scale filamentary molecular cloud, while Class II sources (yellow dots) are dispersively distribution.  Foster et al. (2014) only detected a population low-mass protostars in IRDC G34.43+0.24, and find that the population appears to be distributed along IRDC G34.43+0.24 rather than exclusively associated with the dense clumps. We found that some clustered Class I sources not only coincide with the dense clumps of IRDC G34.43+0.24, but also  with its southern molecular cloud.  Regarding the geometric distribution of the Class I and Class II sources, we can plot the map of star surface density. Because Class II sources are uniform distribution, we only plot the map for Class I sources. This map was obtained by counting all Class I sources with a detection in the 3.6 $\mu$m, 4.5 $\mu$m, 5.8
$\mu$m, and 8.0 $\mu$m bands in squares of $4'\times 4'$, as shown in  Fig. 12 (right panel). From  Fig. 12 (right panel),  we can see that there are clear signs of clustering along the filamentary IRDCs, where IRDC G34.43+0.24 MM1-MM9 (Rathborne et al. 2006), as well as  H {\small II} regions G34.26+0.15 and G34.24+0.13 (Hunter et al. 1998) are located on the peak position of each clustering stars.

\section{DISCUSSIONS}
\subsection{Dynamic Structure of IRDC G34.43+0.24}
From the Spitzer-IRAC 8 $\mu$m emission, IRDC G34.43+0.24 shows a filamentary dark extinction feature. Based on the association with the C$^{18}$O $J$=1-0 emission, the Spitzer-IRAC 8 $\mu$m emission could be divided into regions I, II, and III. Previous ammonia observation indicates that IRDC G34.43+0.24 extends by 9$^{\prime}$ from north to south (Miralles et al. 1994). We found that IRDC G34.43+0.24 should extend by 18$^{\prime}$ from region I to region II. In region III, we  find not only some bright PAHs emission, but also some small IRDCs. The bright PAHs emission is coincident with infrared bubbles B1-B6 (Simpson et al. 2012), and N61 and N62 (Churchwell et al. 2006 \& Deharveng et al. 2010).
From the perspective of large scale, the C$^{18}$O $J$=1-0 emission of 54 to 65 km s$^{-1}$ shows a filamentary structure extended by $\sim$28$^{\prime}$ from region II to region III. The cold dust emission  traced by the ATLASGAL 870 $\mu$m further confirms the existence of the filamentary molecular cloud. We suggest that the previous identified IRDC G34.43+0.24 (9$^{\prime}$) is only part of the larger filamentary cloud.   From the velocity-field map of C$^{18}$O $J$=1-0 for IRDC G34.43+0.24, we conclude that the filament is a single coherent object, which should extend by 34$^{\prime}$ from region I to region III.

Moreover, the molecular clump in region III displays two elongated structures in the $^{13}$CO line along its northwest. Based on the Spitzer-IRAC [4.5]/[3.6] flux ratio, Liu et al. (2013) identified several elongated structures in the same position. They suggested that if the [4.5]/[3.6] ratio traces the distribution of shocked gas, these elongated structures may indicate the multiple jets generated from the G34.26+0.15 complex. Since they also identified the broad line wings in the HCN $J$=3-2 line toward the complex, indicating that there seem to exist the energetic outflow motions. We found that the two elongated structures are well correlated with the 8 $\mu$m dark extinction emission and the 870 $\mu$m cold dust emission (see Fig. 3 and Fig. 1 (c)). Figure 13 shows the position-velocity (PV) diagrams constructed from the $^{13}$CO $J$=1-0 and C$^{18}$O $J$=1-0 emission along the long filamentary IRDC G34.43+0.24. The elongated structures have obvious velocity gradients with respect to the systemic velocity.  The direction of velocity gradient is shown by the red dashed lines. If the elongated structures with the velocity gradient indicate the outflow gas, which is heated, then we cannot detect the dark gas and cold dust emission. Hence, we conclude that the two elongated structures may be two small IRDCs with the high velocity. The observations of more molecular tracer with higher spatial resolutions are needed. For the filamentary IRDC G34.43+0.24, we did not detect the 8 $\mu$m dark extinction emission that connects with its southern part, but identify some small IRDCs and infrared bubbles. If the infrared bubbles emerge, it means that the adjacent IRDC would be segmented and disrupted, and then evolved to the terminal stage (Jackson et al. 2010). From the positions of the small IRDCs, we suggest that the south part of IRDC G34.43+0.24 in region III may be disrupted into some small IRDCs by adjacent star formation activity, such as the outflow motion.

Table 2 lists the name (1) of the infrared bubbles and H {\footnotesize II} regions in the region III of IRDC G34.43+0.24, (2)-(3) equatorial coordinates, (4) velocity ($V_{\rm LSR}$), and (5) the associated velocity ranges.  Because N61 and N62 are associated with H {\footnotesize II} regions 34.325+0.211 and 34.172+0.175, we assume that the bubble and H {\footnotesize II} region are consistent with the same $^{13}$CO emission, respectively.  The $^{13}$CO $J$=1-0 emission in velocity interval 56 to 60 km s$^{-1}$ is associated with H {\footnotesize II} region G34.172+0.175. G34.172+0.175 has the RRL velocity of 62.9 $\pm$ 0.1 km s$^{-1}$ (Anderson et al. 2011), which is not located between this velocity range. Hence, we did not detect the $^{13}$CO $J$=1-0 emission associated with  G34.172+0.175 and N62. Hunter et al. (1998) detected a cool dust core associated with H {\footnotesize II}  region G34.24+0.13, but we did not find the associated $^{13}$CO $J$=1-0 emission. Because the sizes of B4-B6 are smaller, we did not identify the velocity component associated with the $^{13}$CO $J$=1-0 emission.  According to the Galactic rotation model of Fich et al. (1989) together with $R_{\odot}$ = 8.5 kpc and $V_{\odot}$ = 220 km s$^{-1}$, where $V_{\odot}$ is the circular rotation speed of the Galaxy, we derive a kinematic distance of 3.7 kpc based on the average for the whole cloud. Here we will adopt the near kinematic distance. At the distance of 3.7 kpc, the filamentary IRDC G34.43+0.24 has a length of $\sim$37 pc.  Nessie IRDC is  rare object so far. Jackson et al. (2010) obtained that the length of Nessie IRDC is 80 pc, but Goodman et al. (2014) suggested that Nessie is probably longer than 80 pc. A similar filament of more than 50 pc length has been presented by Kainulainen et al. (2011).  We obtained the total mass of the filamentary IRDC G34.43+0.24 are $\sim$4.8$\times10^{4}$$\rm M_{ \odot}$ from region II to region III.  The stability of the filament can be estimated by the virial  parameter $\alpha$=$M_{\rm vir}/M$=$2\sigma_{v}^{2}l/(GM)$ (Fiege \& Pudritz 2000), where $\sigma_{v}$=$\triangle V_{\rm FWHM}/(2\sqrt{2\rm ln2})$  is the average velocity dispersion of C$^{18}$O $J$=1-0, $l$ is the length of the filament, and G is the gravitational constant.  $\triangle V_{\rm FWHM}$ is the mean full width at half-maximum (FWHM) of C$^{18}$O $J$=1-0 emission. In the filament, we found the mean FWHM to be 4.9 km $\rm s^{-1}$, while its length is 30 pc from region II to region III. The mean thermal broadening can be given by $V_{\rm therm}$ =$\sqrt{k T_{\rm ex}/(\mu_{g}m(\rm H_{2})) }$. Region II has an excitation temperature of 10 K, while an excitation temperature of 20 K  is adopted for region III. Adopting $T_{\rm ex}$=20 K, we obtain the maximum $V_{\rm therm}$ $\approx$ 0.3 km s$\rm ^{-1}$, which is less than the mean FWHM of 4.9 km $\rm s^{-1}$. This confirms the supersonic nature of this filament. The virial parameter $\alpha$ is thus estimated to be 1.5, indicating that the filament is stable as a whole, despite infall motion found in  H {\footnotesize II} complex G34.26+0.15.

If the turbulence is dominant, a critical mass to length ratio can be estimated as ($M/l$)$_{\rm crit}$=84$(\Delta V)^{2}M_{\odot}$ $ \rm pc^{-1}$ (Jackson et al. 2010). Then, we obtain ($M/l$) $_{\rm crit}$ $\approx$ 2.0$\times$10$^{3}$ $M_{\odot}$ $ \rm pc^{-1}$. Besides, using the total mass ($\sim$4.8$\times10^{4}$$\rm M_{ \odot}$)  and length ($\sim$37 pc) of the filament from region II to region III, we derived a linear mass density $M/l$ of $\sim$ 1.6$\times$10$^{3}$ $M_{\odot}$ $ \rm pc^{-1}$, which is roughly consistent with ($M/l$) $_{\rm crit}$.  Considering the uncertainties of the FWHM, the turbulent motion may be helping stabilize the filament against the radial collapse.

\subsection{Star Formation Scenario}
IRDCs have been proposed to be the birthplace of massive stars and their host clusters (Carey et al. 2000; Egan et al. 1998; Rathborne et al. 2006). Rathborne et al. (2006) identified nine clumps in IRDC G34.43+0.24. Three of these cores exist the massive star formation activity, the remaining cores are the starless cores. In region III, UC H II region G34.24+0.13 and H II complex G34.26+0.15 are embedded in a filamentary molecular clump.  H II complex 34.26+0.15 contains four small H {\footnotesize II} region (Reid \& Ho 1985). The cometary G34.26+0.15C is not only associated with a hot molecular core (Hunter et al. 1998; Campbell et al. 2004; Mookerjea et al. 2007), but also has an infall motion (Liu et al. 2013), indicating that UC H {\footnotesize II} complex G34.26+0.15 is an ongoing massive star formation region. Additionally, several infrared bubbles are located in region III.  The bubbles can be produced by stellar wind and overpressure by ionization and heating by stellar UV radiation from O and early-B stars. The infrared bubbles emerge, suggesting that several massive stars have formed in this region. Especially, N61 has a double-shell structure, the outer traced by 8 $\mu$m and $^{13}$CO $J$=1-0 emission and the inner traced by 24 $\mu$m and 21 cm emission. The outer shell (9.9$\times10^{5}$ yr) is created by the expansion of H II region G34.172+0.175 in the filamentary molecular clump, while the inner shell (4.1$\sim$6.3$\times$10$^{5}$ yr) may be produced by the energetic stellar wind from its central massive star. We determined that the spectral type of the ionizing star of G34.172+0.175 is between B0.5V and B1V. Since the timescales of the double shells are much less than the main-sequence  lifetime of such stars (1.3$\sim$1.6$\times$10$^{7}$ yr for B1V$\sim$B0.5V stars, Chen et al. 2013), we infer that the central massive star of N61 is still in the main-sequence stage. To summarize, region I shows a filamentary dark extinction, but no star formation activity are detected; Region II with IRDC G34.43+00.24 presents a filamentary dark extinction associated with C$^{18}$O emission. A UC H II region and three massive protostars are identified in this region; Region III shows a filamentary C$^{18}$O molecular cloud, but no filamentary dark extinction is detected except for some small IRDCs. Moreover, some massive protostars, UC H II regions, and even infrared bubbles are detected in region III.  We suggest that IRDC G34.43+0.24 is in different evolutionary stages from region I to region III.  The star-forming filamentary clouds represent a later evolutionary stage in the life of an IRDC (Jackson et al. 2010). IRDC G34.43+0.24 in region III appear in a later evolutionary stage such as Orion and NGC 6334 (Kraemer et al. 1999).

Furthermore, Shepherd et al. (2007) identified some massive protostars with an age of about 10$^{5}$ yr after they detected a low-mass population of stars ($\sim$ 10$^{6}$ yr) around IRDC G34.43+0.24 MM2 (Shepherd et al. 2004). They suggested that the stars in this region may have formed in two stages: first lower mass stars formed and then more massive stars began to form. Because Foster et al. (2014) detected a population low-mass protostars that are distributed along filamentary IRDC G34.43+0.24 rather than exclusively associated with the dense clumps, they concluded that massive stars predominantly from in the most bound parts of the filaments, while the low-mass protostars form in less bound portions of the filament. From the GLIMPSE I catalog, we selected some Class I sources with an age of $\sim$$10^{5}$ yr on the basis of infrared color indices. It is interesting note that these Class I sources are clustered distribution along the filamentary molecular cloud. The north star cluster is associated with IRDC G34.43+0.24, while the south star cluster is well correlated with the filamentary molecular clump in region III. In addition, IRDC G34.43+0.24 MM1-MM9 (Rathborne et al. 2006), as well as  H {\small II} regions G34.26+0.15 and G34.24+0.13 (Hunter et al. 1998) are located on the peak position of each clustering stars.  These results are consistent with previous suggestions that IRDCs are the precursors of massive stars and star clusters. If these selected Class I sources are most low-mass protostars as the observations of Foster et al. (2014), the low-mass protostars may form contemporaneously with high-mass protostars in such a filament. Based on the presence of distributed low-mass star formation, Foster et al. (2014) suggest that the low-mass and massive protostars certainly may form by the ``sausage'' instability in IRDC G34.43+0.24. The ``sausage'' instability model in IRDCs is proposed by Jackson et al. (2010). Comparing the age of H {\footnotesize II} region  G34.172+0.175 (9.9$\times10^{5}$ yr) with that of Class I sources ($\sim$10$^{5}$ yr), we infer that the ionizing stars of G34.172+0.175 may be the first generation of massive stars that formed in such a filament.

\section{CONCLUSIONS}
We present the molecular $^{13}$CO $J$=1-0 and C$^{18}$O $J$=1-0, infrared, and  radio continuum observations toward IRDC G34.43+0.24 ob a large scale.  The main results are summarized as follows:

1. At 8 $\mu$m (Spitzer - IRAC), IRDC G34.43+0.24 appears to be a dark filament along the north-south direction. Both the ATLASGAL 870 $\mu$m and C$^{18}$O $J$=1-0 images exhibit bright filamentary structures containing massive protostars, UC H II regions, and infrared bubbles.  The spatial extend of IRDC G34.43+0.24 is about 37 pc (34$^{\prime}$) at a distance of 3.7 kpc.  IRDC G34.43+0.24 has a linear mass density of $\sim$ 1.6$\times$10$^{3}$ $M_{\odot}$ $ \rm pc^{-1}$, which is roughly consistent with its critical mass to length ratio, suggesting that the turbulent motion may be helping stabilize the filament against the radial collapse.

2. IRDC G34.43+0.24 could be divided into three portions. Each portion may be in different evolutionary stage. The southern portion of IRDC G34.43+0.24 containing the strong star-forming activity and several smaller IRDCs, may evolute a later stage in the life of an IRDC.

3. Infrared bubble N61 has expanded into IRDC G34.43+0.24, which has a double-shell structure, the outer traced by 8 $\mu$m and $^{13}$CO $J$=1-0 emission and the inner traced by 24 $\mu$m and 20 cm emission.  N61 is associated with H {\small II} region G34.172+0.175. We conclude that the outer shell with a timescale of 9.9$\times10^{5}$ yr is created by the expansion of H {\small II} region G34.172+0.175, while the inner shell with an age of 4.1$\sim$6.3$\times$10$^{5}$ yr may be produced by the energetic stellar wind from its central massive star. N62 shows a ringlike structure at 8 $\mu$m, which is associated with H {\small II} region G34.325+0.211 with an age of (5.6$\pm$0.1)$\times10^{5}$ yr.

4. The selected Class I sources with an age of $\sim$$10^{5}$ yr are clustered along the filamentary molecular cloud. The north star cluster is associated with IRDC G34.43+00.24, while the south star cluster is well correlated with the filamentary molecular clump. IRDC G34.43+0.24 MM1-MM9, as well as  H {\small II} regions G34.26+0.15 and G34.24+0.13 are located on the peak positions of each clustering stars. Comparing the age of H {\footnotesize II} region  G34.172+0.175 (9.9$\times10^{5}$ yr) with that of Class I sources ($\sim$10$^{5}$ yr), we infer that the ionizing stars of G34.172+0.175 may be the first generation of massive stars that formed in IRDC G34.43+0.24.

\acknowledgments We are very grateful to the anonymous referee for his/her helpful comments and suggestions. This work is made use of data from the Spitzer Space Telescope, which is operated by the Jet Propulsion Laboratory, California Institute of Technology under a contract with NASA. The ATLASGAL project is a collaboration between the Max-Planck-Gesellschaft, the European Southern Observatory (ESO) and the Universidad de Chile. It includes projects E-181.C-0885, E-078.F-9040(A), M-079.C-9501(A), M-081.C-9501(A) plus Chilean data.  We are also grateful to the staff at the Qinghai Station of PMO for their assistance during the observations. Thanks for the Key Laboratory for Radio Astronomy, CAS to partly support the telescope operating. This work was supported by the National Natural Science Foundation of China (Grant No. 11363004 and 11403042).  This paper is also partly supported by National Key Basic Research Program of China(973 Program) 2015CB857100. DL acknowledges the support from the Guizhou Scientific Collaboration Program (20130421).

\clearpage
\begin{figure}
\vspace{-3mm}
\includegraphics[angle=270,scale=0.7]{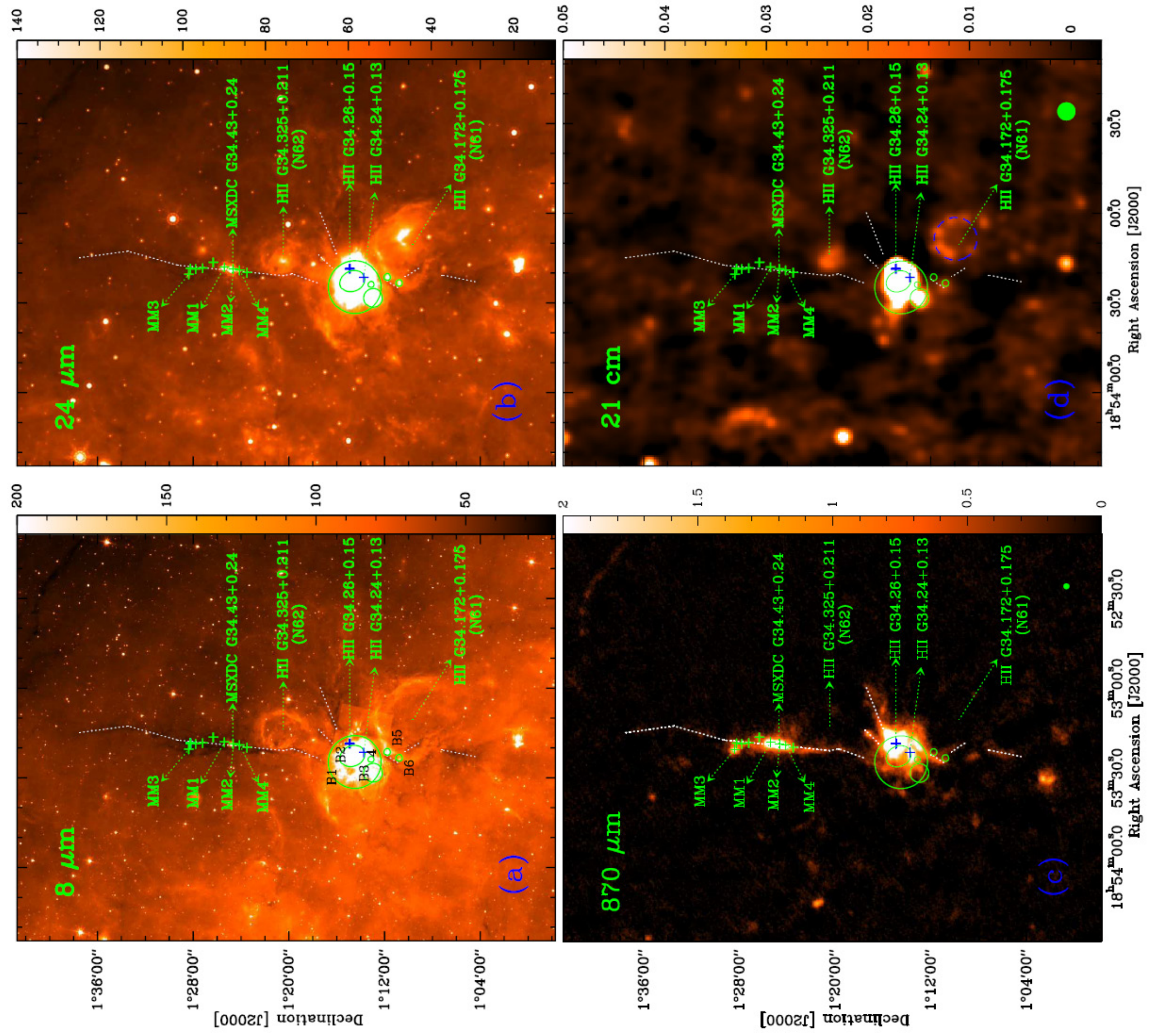}
\vspace{-3mm}\caption{(a) the Spitzer-IRAC 8
$\mu$m emission.  (b) the Spitzer-MIPSGAL 24 $\mu$m emission. (c) the ATLASGAL 870 $\mu$m emission. (d) the NVSS 21 cm emission. The unit of color bar is MJy sr$^{-1}$ for 8 and 24$\mu$m, while Jy beam$^{-1}$ for 870 $\mu$m and 21 cm.
In each panel, the white dashed lines mark the IRDCs. The nine green pluses show the positions of IRDC G34.43+0.24 MM1-MM9 (Rathborne et al. 2006). The two blue pluses present the positions of H {\small II} regions G34.26+0.15 and G34.24+0.13 (Hunter et al. 1998). The six green circles represent the bubbles identified by Simpson et al. (2012). }
\end{figure}

\begin{figure*}
\vspace{-7mm}
\includegraphics[angle=270,scale=0.62]{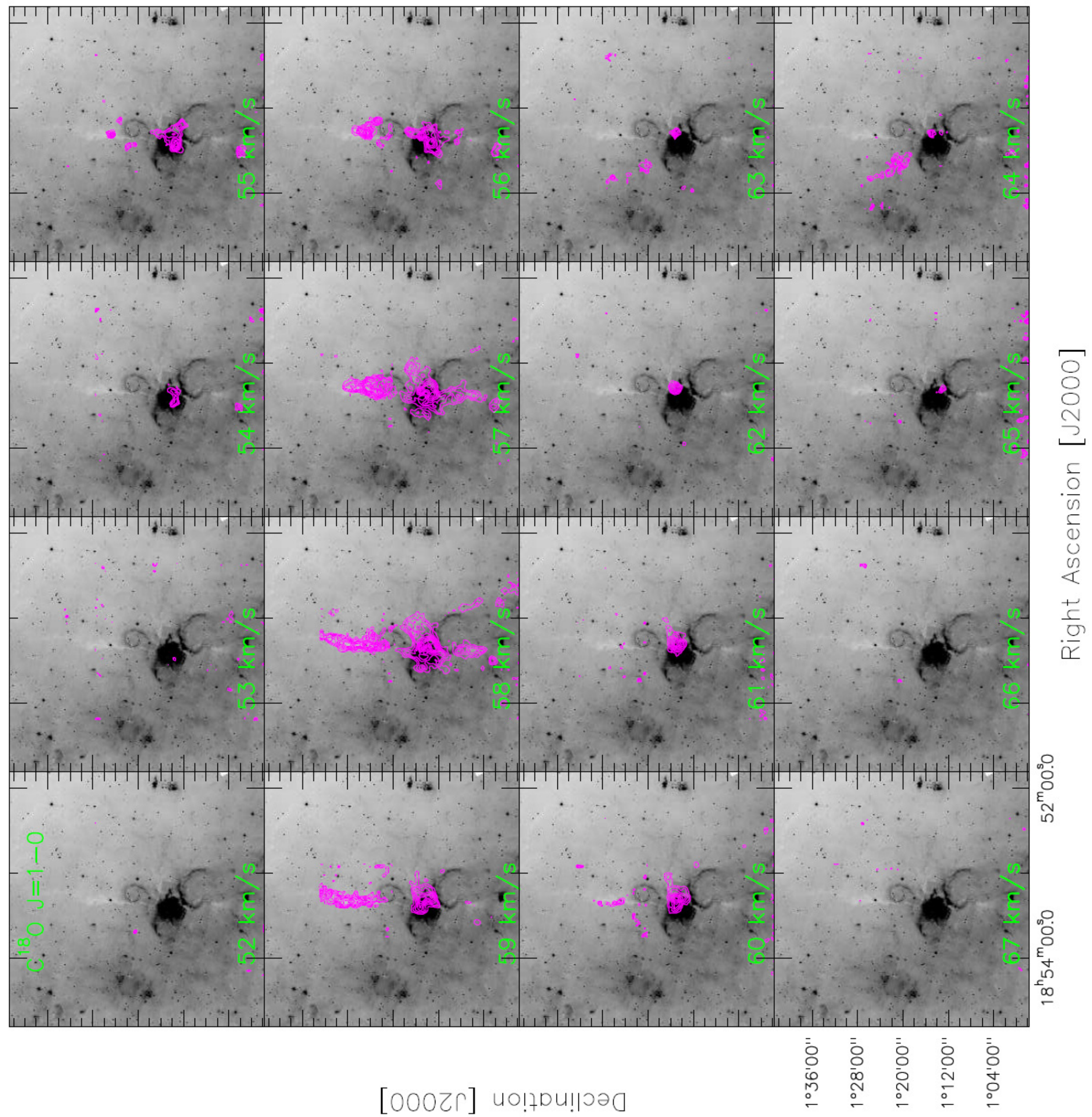}
\vspace{-3mm}\caption{C$^{18}$O $J$=1-0 channel maps in step of 1 km $s^{-1}$ overlaid on the Spitzer-IRAC 8
$\mu$m emission (grey). Central velocities are indicated in each image.}
\end{figure*}

\begin{figure}[]
\vspace{0mm}
\includegraphics[angle=0,scale=0.55]{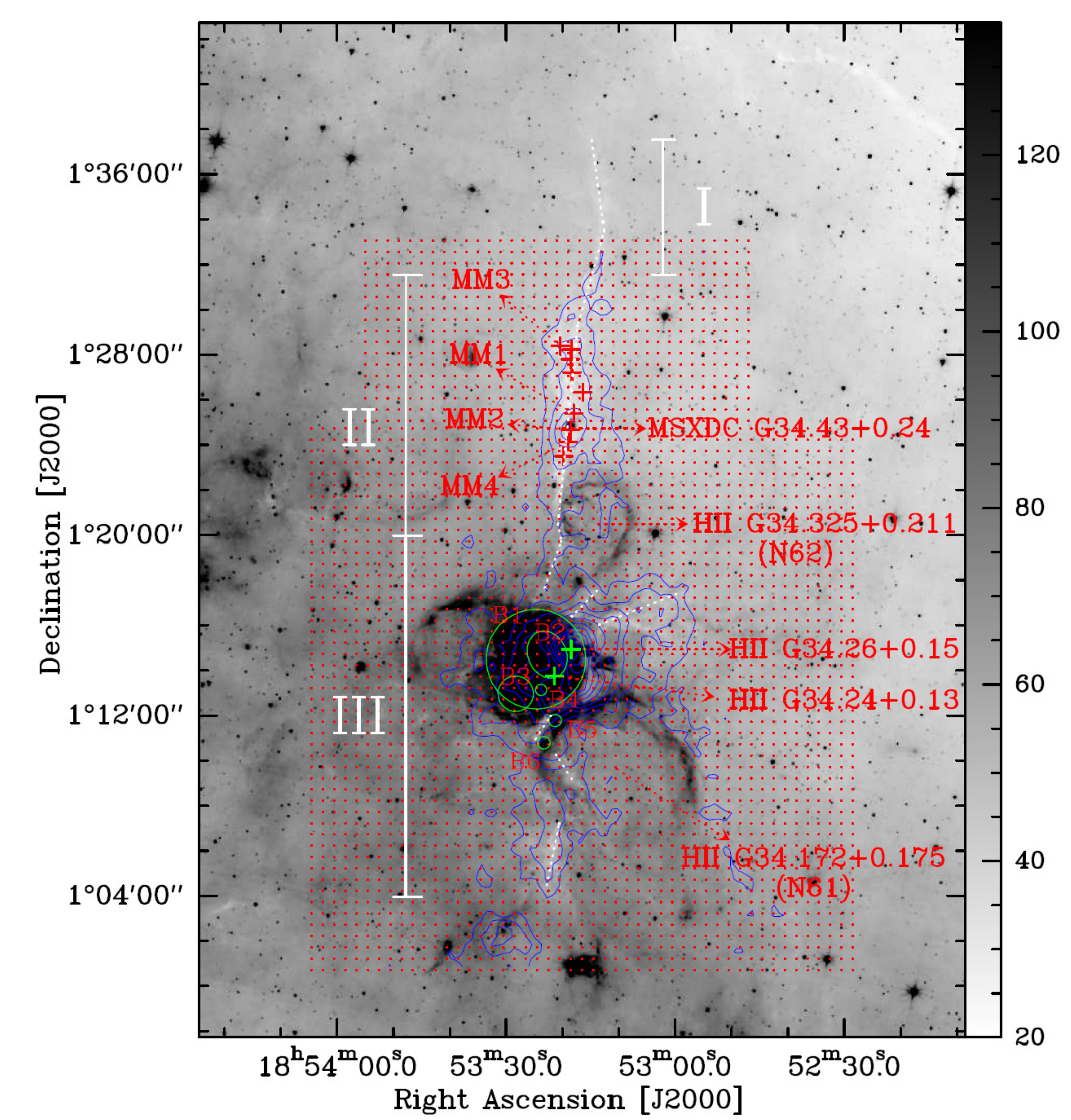}
\vspace{-3mm}\caption{Contours of C$^{18}$O $J$=1-0 emission superimposed on the 8
$\mu$m emission map (grey). The integrated velocity is from 53 to 65 km s$^{-1}$. The contour
levels are from 2.8 to 23.1  by a step of 1.7 K km s$^{-1}$ . The six {\bf green} circles show the bubbles identified by Simpson et al. (2012). I, II, and III may represent the evolutive stage of IRDC.  The {\bf red }dot symbols mark the mapping
points.  The right color bar is the unit in MJy sr$^{-1}$. }
\end{figure}

\begin{figure}[]
\vspace{-6mm}
\includegraphics[angle=0,scale=.395]{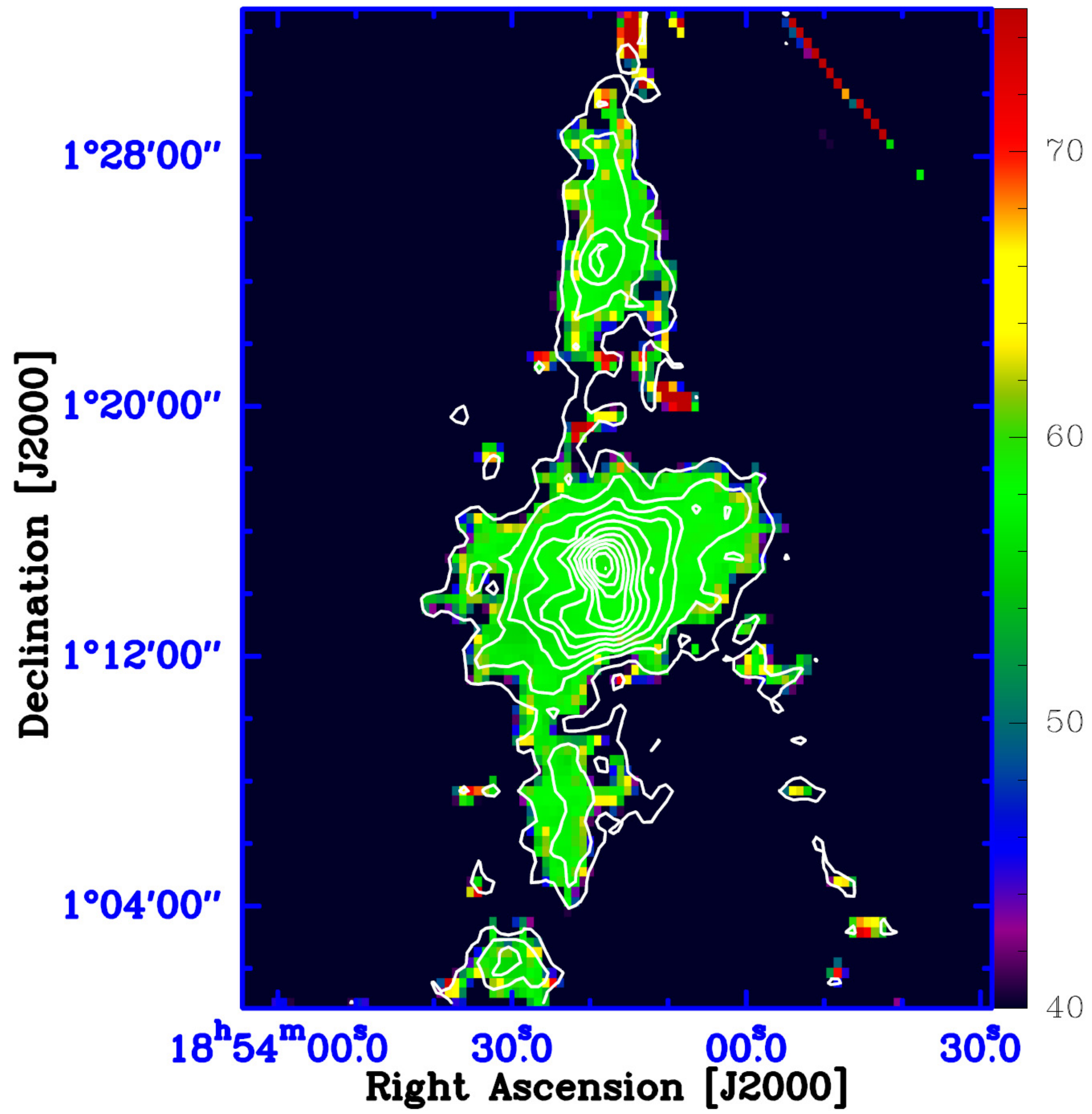}
\vspace{-7mm}\caption{Velocity-field (moment 1) map of C$^{18}$O $J$=1-0 overlaid with its integrated intensity contours (white).}
\end{figure}

\begin{figure*}
\vspace{0mm}
\includegraphics[angle=90,scale=0.55]{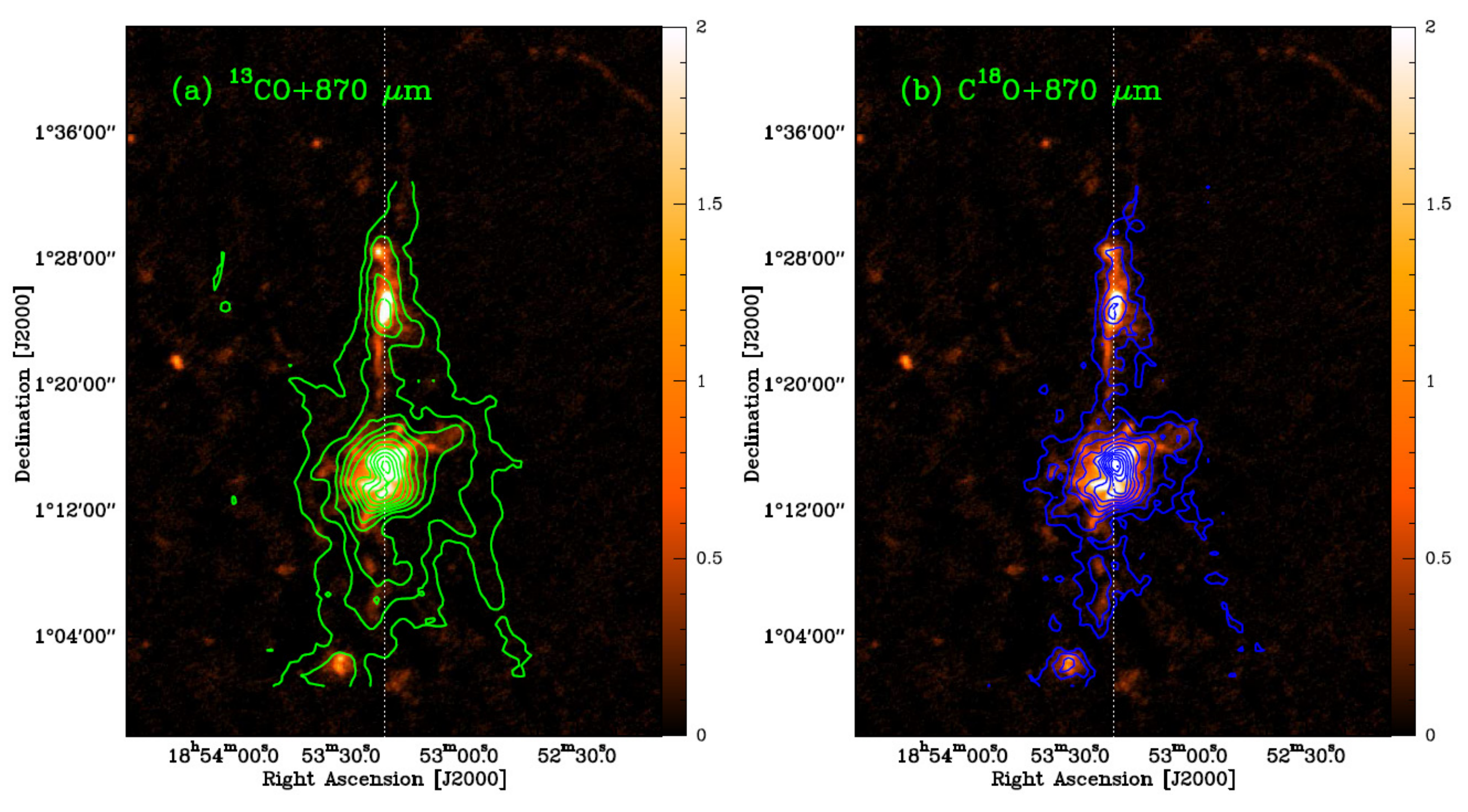}
\vspace{-5mm}\caption{(a) $^{13}$CO $J$=1-0 integrated intensity map overlaid on the  ATLASGAL 870 $\mu$m image (color scale). The green contour levels are from 12.3 (5$\sigma$) to 77.6  by a step of 5.9 K km s$^{-1}$ . (b) C$^{18}$O $J$=1-0 integrated intensity map overlaid on  ATLASGAL the 870 $\mu$m image (color scale). The blue contour
levels are from 2.8 (5$\sigma$) to 23.1  by a step of 1.7 K km s$^{-1}$ . The integrated velocities of both $^{13}$CO $J$=1-0 and C$^{18}$O $J$=1-0 are from 53 to 65 km s$^{-1}$. The white dashed lines show the direction of the PV diagrams (See Figure 5).}
\end{figure*}

\begin{table*}[]
\begin{center}
\tabcolsep 5.8mm \caption{The physical parameters of region II and region III.}
\def\temptablewidth{10\textwidth}
\begin{tabular}{ccccccccc}
\hline\hline\noalign{\smallskip}
Name   & Trace& Area &$N_{\rm H_{2}}$ & $n(\rm H_{2})$& $M$\\
       &      &  (arcmin$^{2}$)  &(cm$^{-2}$) &(cm$^{-3}$)  & ($\rm M_{ \odot}$)  \\
  \hline\noalign{\smallskip}
Region II     &C$^{18}$O&  33.6  &2.4$\times10^{22}$  & 5.5$\times10^{2}$ & 6.0$\times10^{3}$ \\  
Region III    &C$^{18}$O&       109.3  &4.5$\times10^{22}$  & 5.7$\times10^{2}$ & 4.2$\times10^{4}$ \\  
\noalign{\smallskip}\hline
\end{tabular}\end{center}
\end{table*}

\begin{figure*}
\vspace{0mm}
\includegraphics[angle=270,scale=0.8]{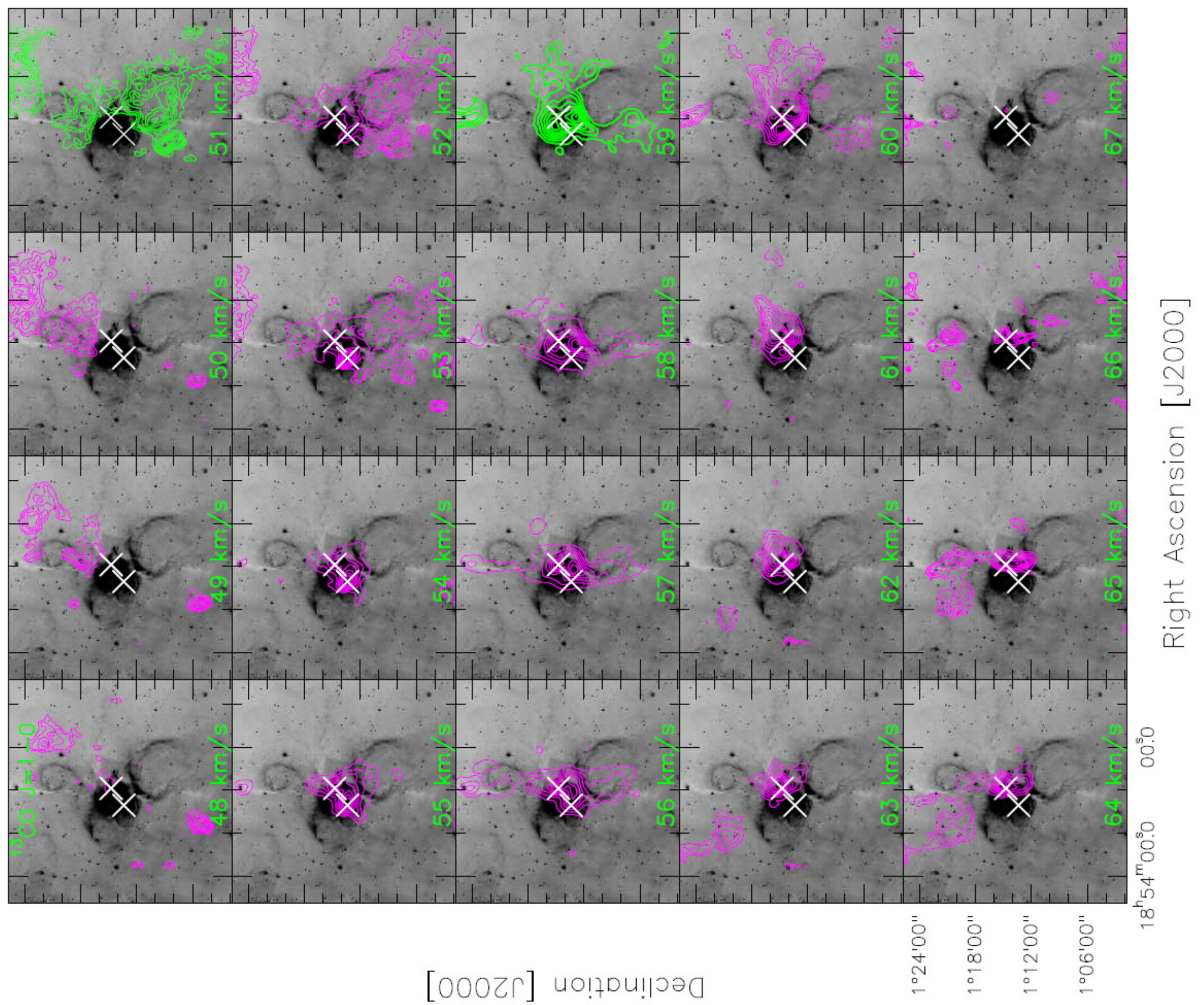}
\vspace{-5mm}\caption{$^{13}$CO $J$=1-0 channel maps (pink and green contours) in step of 1 km $s^{-1}$  overlayed on the Spitzer-IRAC 8 $\mu$m emission (grey). Central velocities are indicated in
each image. The two red crosses mark the peak positions of the identified cores. }
\end{figure*}

\begin{table}
\begin{center}
\caption{The parameters of associated sources with IRDC G34.43+0.24
 \label{tbl-1}}
\begin{tabular}{rcccccccccr}
\tableline\tableline
Source name & R.A. & Decl. & $V_{\rm LSR}$ &  Velocity range    \\
     & (J2000.)     & (J2000.)      & km s$^{-1}$ &  km s$^{-1}$     \\
\tableline
\multicolumn{6}{c}{IRDC \& H {\footnotesize II} regions}    \\
\tableline
IRDC G34.43+0.24                          &18 53 18.9   & 01 26 38.6   &57.6$\pm$0.1$^{a}$ &(53.0--65.0)  \\
H{\footnotesize II} G34.325+0.211         &18 53 13.8   & 01 20 08.0   &62.9$\pm$0.1$^{b}$ &(56.0--60.0)    \\
H{\footnotesize II} G34.26+0.15           & 18 53 20.1  & 01 14 37.1   &54.6$^{c}$ &(53.0--66.0)    \\
H{\footnotesize II} G34.24+0.13           & 18 53 21.5  & 01 13 45.3   &57.0$^{d}$ &--             \\
H{\footnotesize II} G34.172+0.175         & 18 53 04.7  & 01 11 02.0   &57.3$\pm$0.1$^{b}$ &-- &   \\
\tableline
\multicolumn{6}{c}{Infrared bubbles}    \\
\tableline
N61                 & 18 53 10.5 & 01 09 14.9  & -- & (56.0--60.0)   \\
N62                 & 18 53 13.6 & 01 20 46.7  &      --            & --   \\
{\footnotesize MWP G034246+001023(B1)}  & 18 53 28.3 & 01 12 58     & 51.0$\pm$0.5 & (48.0--53.0)   \\
{\footnotesize MWP G034261+001357(B2)}  & 18 53 22.7 & 01 14 42     &--    & (52.0--56.0)  \\
{\footnotesize MWP G034262+001267(B3)}  & 18 53 24.7 & 01 14 30     &--    & (56.0--60.0)   \\
{\footnotesize MWP G034240+001200s(B4)} & 18 53 23.8 & 01 13 08     &--    & --              \\
{\footnotesize MWP G034210+001200s(B5)} & 18 53 20.5 & 01 11 32     &--    &--              \\
{\footnotesize MWP G034200+001000s(B6)} & 18 53 23.7 & 01 10 27     &--    & --             \\
\tableline
\end{tabular}
\end{center}
Notes.(a) Miralles et al. 1994; The NH$_{3}$ line. (b) Anderson et al. 2011; The radio recombination line.
(c) Kolpak et al. 2003; The radio recombination line. (d) Hunter et al. 1998; The H$_{2}$CO line.
\end{table}

\begin{figure*}
\vspace{-7mm}
\includegraphics[angle=0,scale=0.55]{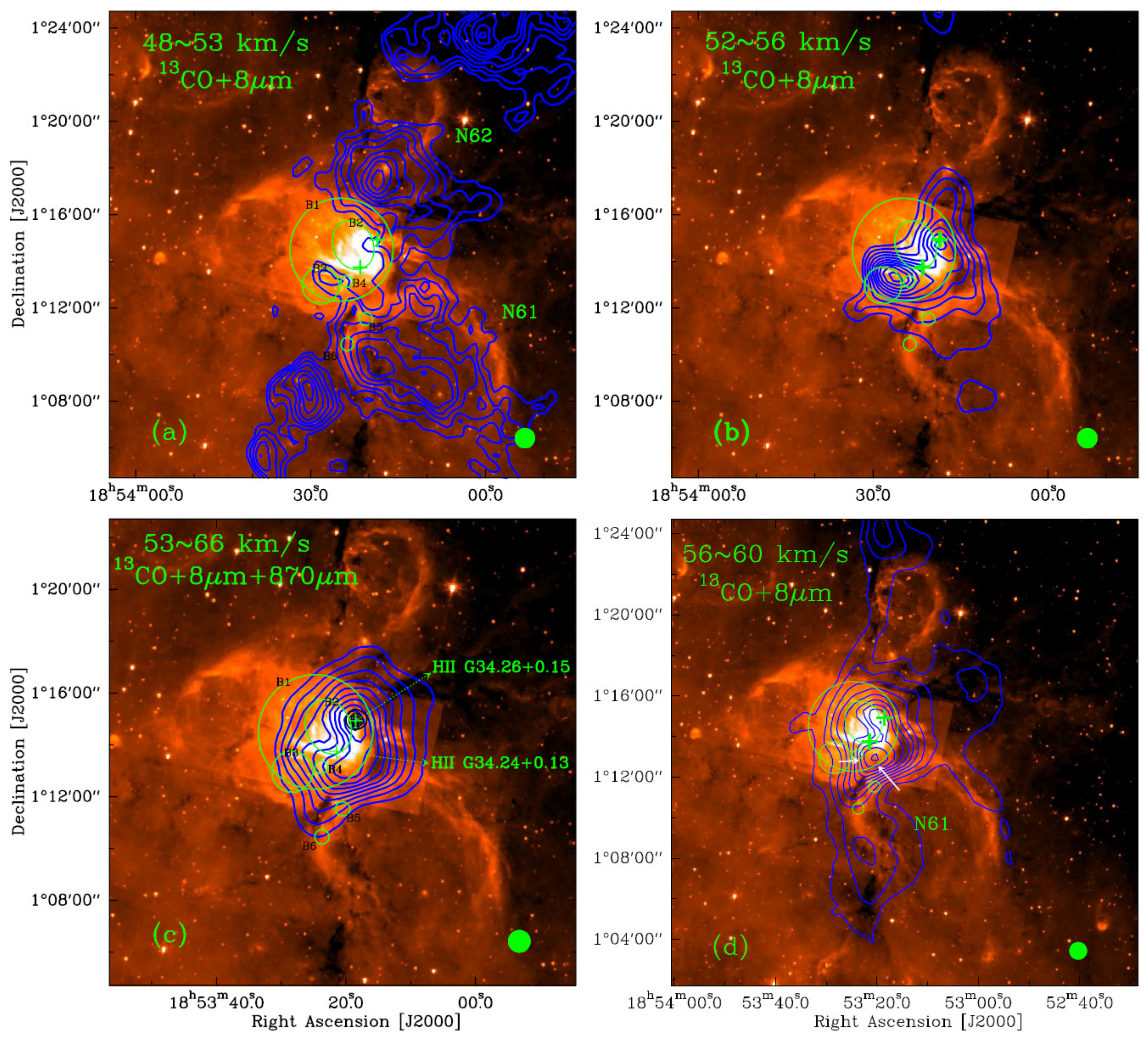}
\vspace{-5mm}\caption{Panels (a), (b), and (d) show the $^{13}$CO $J$=1-0 emission (green contours) overlayed on the 8
$\mu$m emission (heat scale) maps , while panel (c) shows the $^{13}$CO $J$=1-0 emission (green contours) and 870 $\mu$m  emission (black contours) overlayed on the 8 $\mu$m emission (heat scale) maps. The integrated velocities are indicated in the left corner of each panel, respectively. The blue contour levels are 35, 42,... , $98\%$ of eack peak value. The black contour levels (ATLASGAL 870 $\mu$m) in panel (c) are 10.8, 21.7, 32.5, 43.4, and 54.3 Jy/beam. The two  while arrows in panel (d) mark  the direction of velocity gradient. The two green pluses present the positions of H {\small II} regions G34.26+0.15 and G34.24+0.13 (Hunter et al. 1998). The PMO beam is showed in right corner of each panel. }
\end{figure*}

\begin{figure}[]
\vspace{-6mm}
\includegraphics[angle=270,scale=.5]{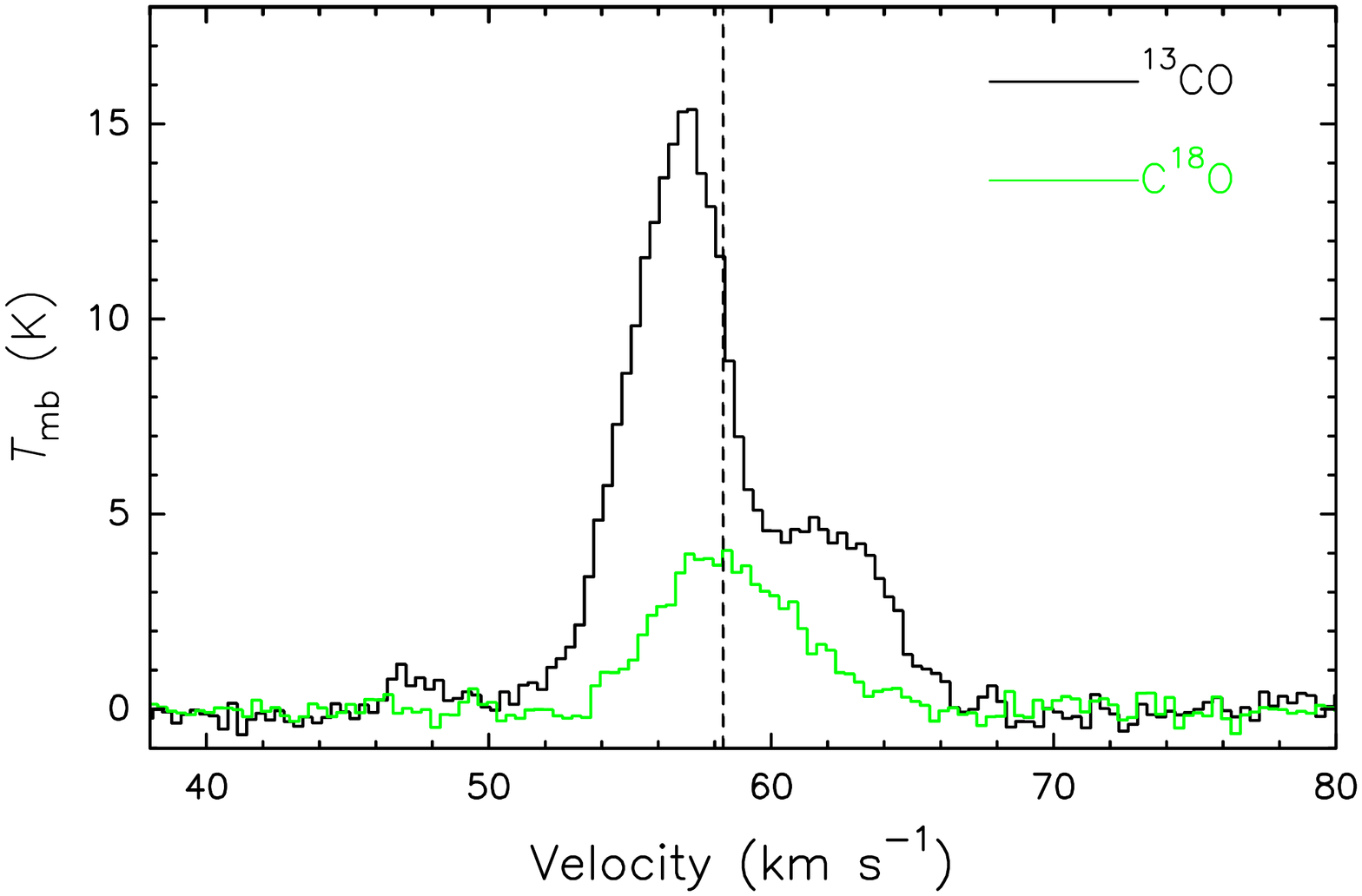}
\vspace{-7mm}\caption{$^{13}$CO $J$=1-0 and C$^{18}$O $J$=1-0 spectra at the peak position of G34.26+0.15 complex.  }
\end{figure}

\begin{figure}[]
\vspace{-6mm}
\includegraphics[angle=0,scale=.52]{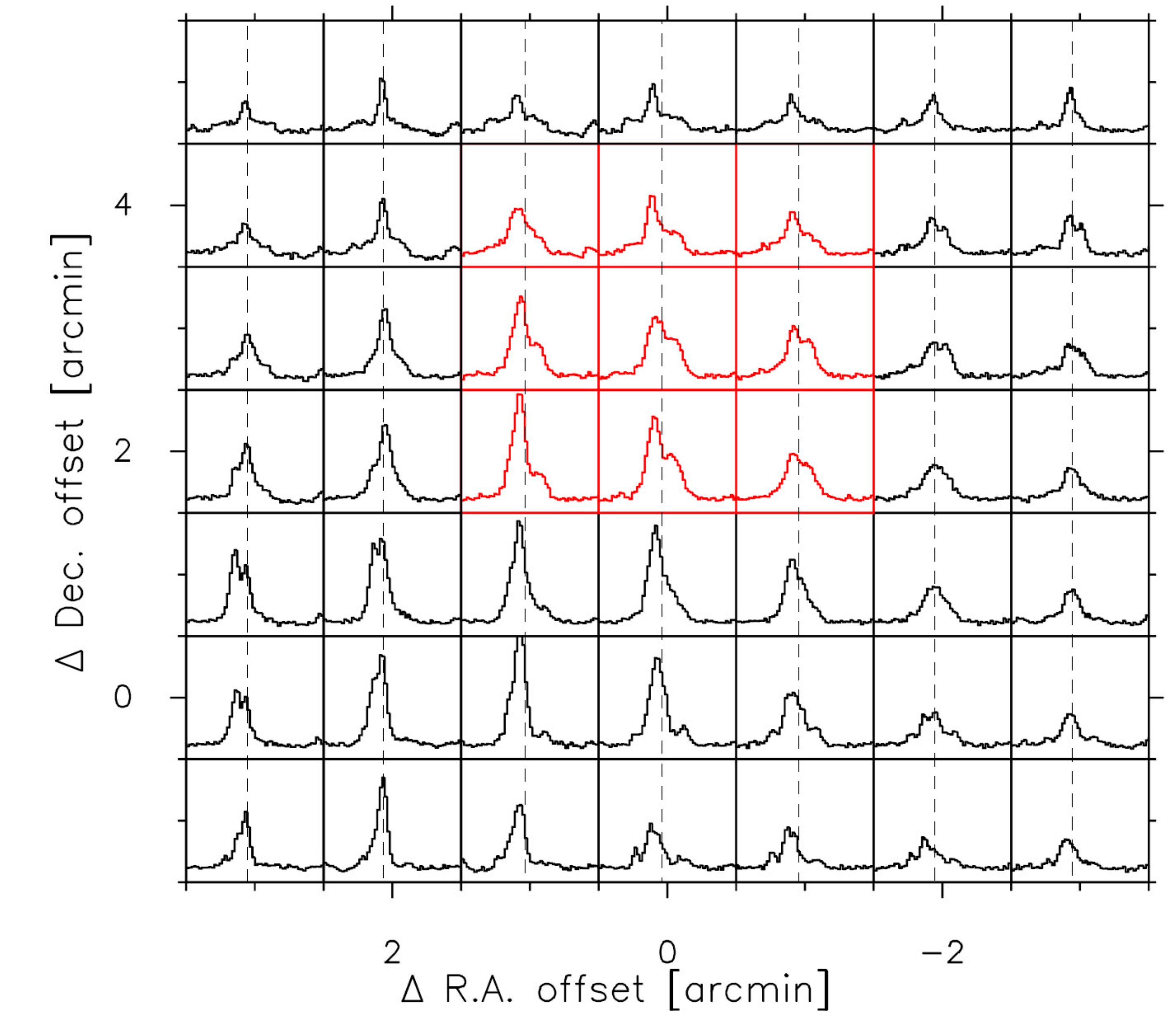}
\vspace{-4mm}\caption{ The mapping grids of G34.26+0.15 complex. The red spectra clearly show the blue-profile signature. Offset(0,0) position is R.A(J2000.0)= $\rm 18 ^{h}53^{m}16^{s}.52$ and
decl.(J2000.0) = $\rm 01 ^{\circ}12^{\prime}43^{\prime\prime}.3$. The black dashed line marks the LSR velocity of 58.2 km s$^{-1}$.}
\end{figure}

\begin{figure}[]
\vspace{0mm}
\includegraphics[angle=0,scale=0.52]{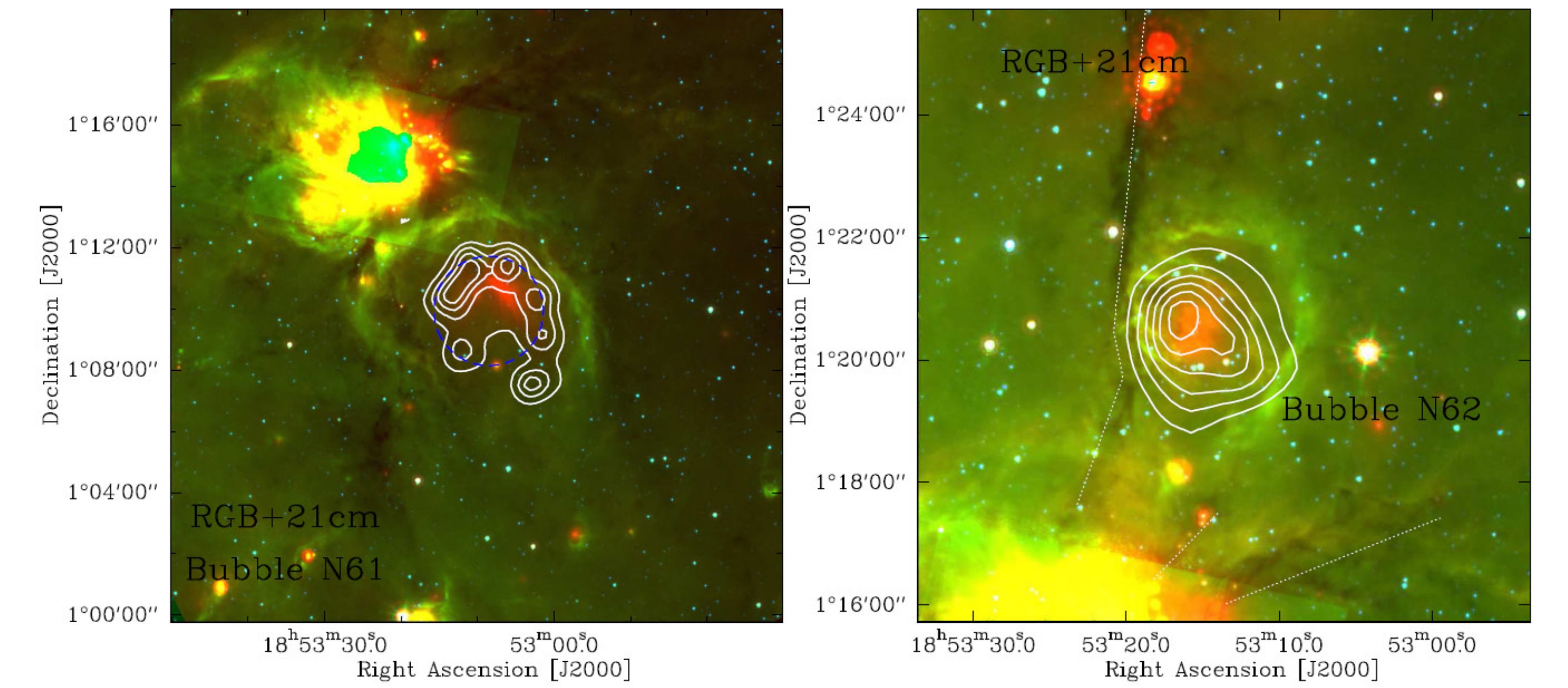}
\vspace{-6mm}\caption{1.4 GHz radio continuum emission
contours (white) overlayed on the three color image of the bubble N61 and N62 composed from the Spitzer 3.6 $\mu$m, 8 $\mu$m, and 24 $\mu$m bands in blue, green, and red, respectively. The blue circle indicates that the ionized gas of H {\footnotesize II} G34.325+0.211 has a central hole.}
\end{figure}

\begin{figure}[h]
\includegraphics[angle=0,scale=.650]{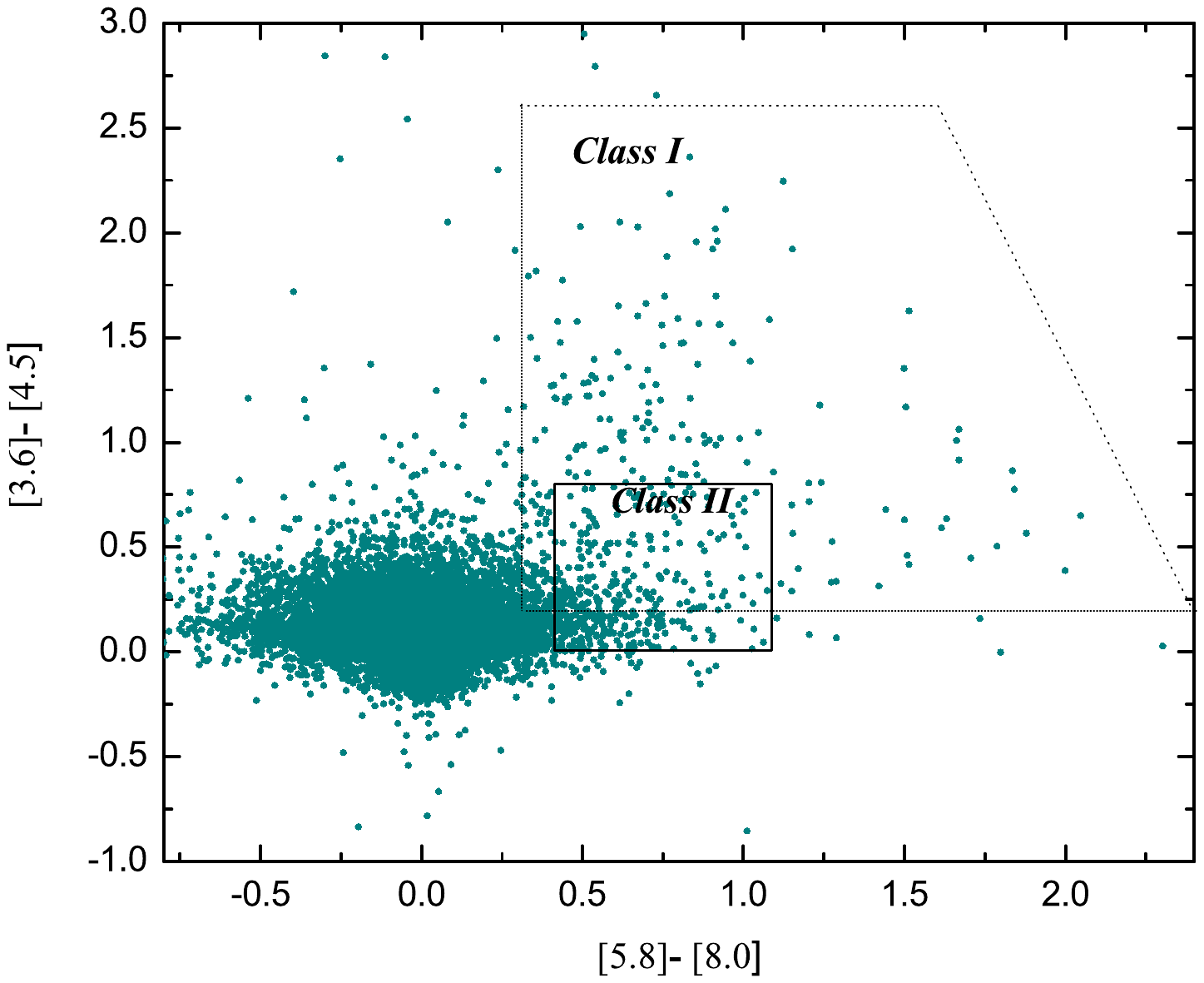}
\vspace{-18mm} \caption{GLIMPSE color--color diagram [5.8]--[8.0] versus [3.6]--[4.5]
for sources. The regions indicate the stellar evolutionary stage as
defined by Allen et al. (2004). Class I sources are protostars with
circumstellar envelopes and Class II are disk-dominated objects.}
\end{figure}

\begin{figure}[h]
\vspace{0mm}
\includegraphics[angle=0,scale=0.65]{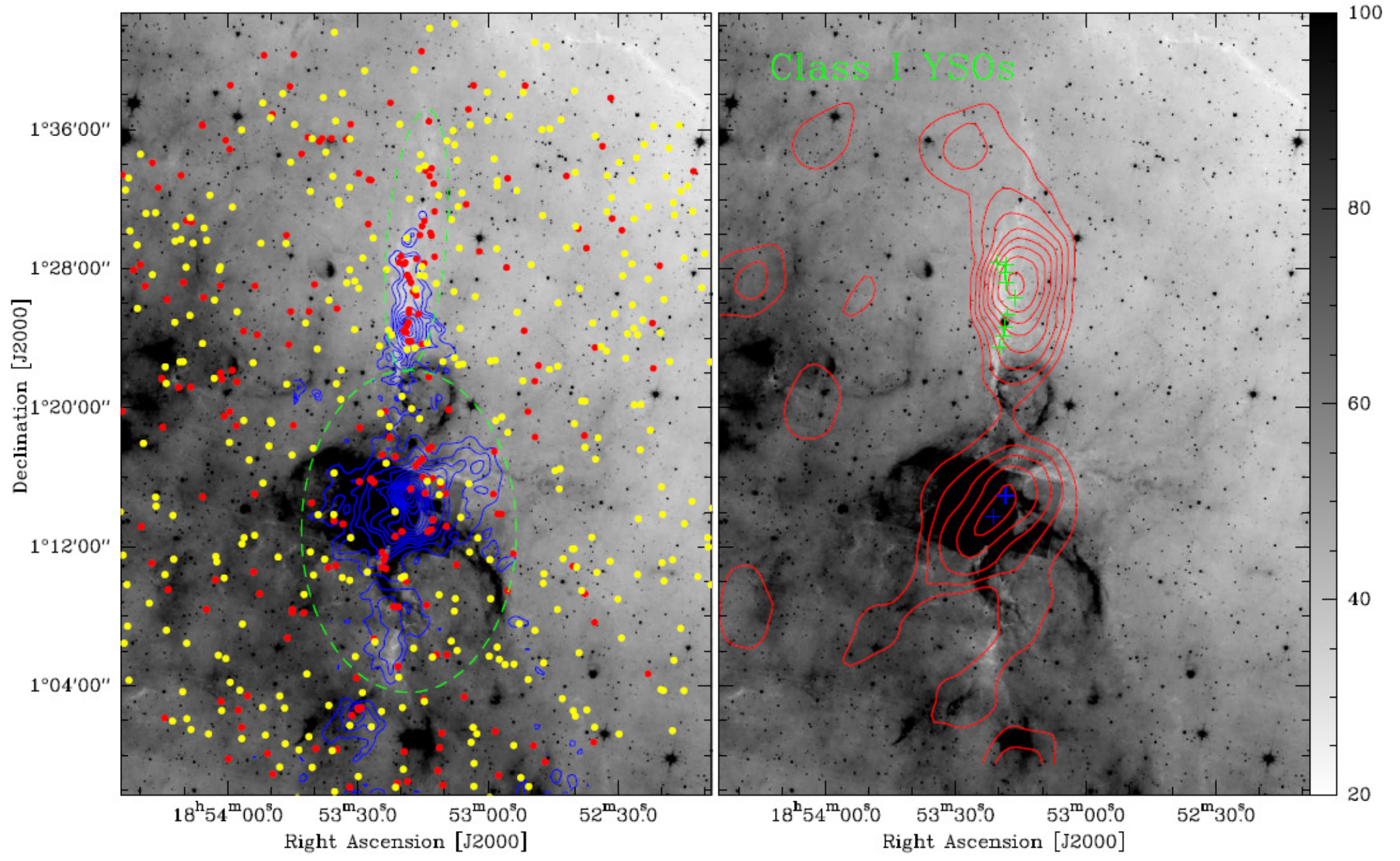}
\vspace{-5mm}
\caption{Left panel: Positions of  Class I and II  sources relative to the
$^{13}$CO $J=1-0$ emission (blue contours) overlaid on the 8 $\mu$m emission. The Class I
sources are labeled as the red dots, and the Class II sources as the yellow  dots. Two dashed ellipses mark the YSOs associated with IRDC G34.43+0.24. Right panel: Stellar-surface density map (red contours) of Class I candidates are superimposed on the 8 $\mu$m emission. Contours range from 4 to 15 stars (4arcmin)$^{-2}$ in steps of 2 stars (4arcmin$)^{-2}$. 1$\sigma$
is 1.3 (4arcmin)$^{-2}$  (background stars). The nine green pluses show the positions of IRDC G34.43+0.24 MM1-MM9 (Rathborne et al. 2006), while
the two blue pluses present the positions of H {\small II} regions G34.26+0.15 and G34.24+0.13 (Hunter et al. 1998), also shown in Fig. 1. }
\end{figure}

\begin{figure*}
\vspace{0mm}
\includegraphics[angle=0,scale=0.55]{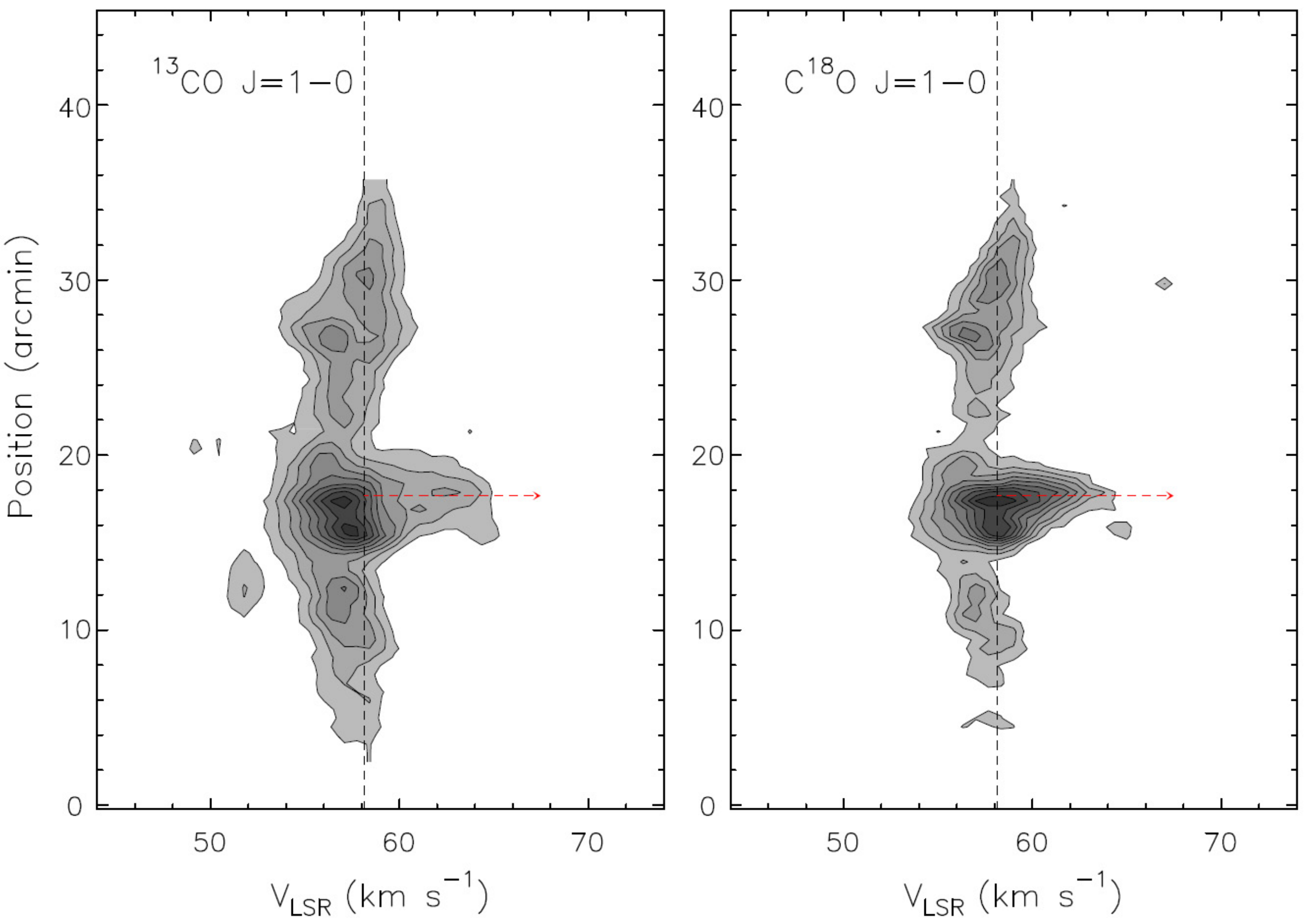}
\vspace{-6mm}\caption{Position-Velocity diagrams of the $^{13}$CO ($J$=1-0) and C$^{18}$O ($J$=1-0) emission along the filamentary IRDC G34.43+0.24 (see the long dashed lines in Figure 4). The position is measured along the long arrow (RA(J2000)=$19^{\rm h}53^{\rm m}16.63^{\rm s}$, DEC(J2000)=$00^{\circ}57'33.61^{\prime\prime}$ to RA(J2000)=$19^{\rm h}53^{\rm m}16.63^{\rm s}$, DEC(J2000)=$01^{\circ}42'38.61^{\prime\prime}$) with a width of 35$^{\prime}$. The black dashed line marks the LSR velocity of 58.2 km s$^{-1}$. The red dashed arrows show the high-velocity gas.}
\end{figure*}


\begin{thebibliography}{}
\bibitem[Allen(2004)]{Allen04} Allen, L. E., Calvet, \& N., D'Alessio, P., 2004, \apjs, 154, 363
\bibitem[Andre(1994)]{Andre94} Andr\'{e}, P., \&  Montmerle, T. 1994, \apj, 420, 837
\bibitem[Anderson(2011)]{Anderson11} Anderson, L. D., Bania, T. M., Balser, D. S., \& Rood, R. T. 2011, \apj, 194, 32
\bibitem[Beuther(2012)]{Beuther12}Beuther, H., Tackenberg, J., Linz, H., et al. 2012, \apj, 747, 43
\bibitem[Beuther(2013)]{Beuther13} Beuther, H., Linz, H., Tackenberg, J., et al. A\&A, 553, 115
\bibitem[Benjamin(2003)]{Benjamin03} Benjamin, R. A., Churchwell, E., \&  Babler, B. L., 2003, PASP, 115, 953
\bibitem[Bronfman(1996)]{Bronfman96} Bronfman, L., Nyman, L.-A., \& May, J. 1996, A\&AS, 115, 81
\bibitem[Campbell(2004)]{Campbell04} Campbell, M. F., Harvey, P. M., Lester, D. F., \& Clark, D. M. 2004, \apj, 600, 254
\bibitem[Carey(2000)]{Carey00} Carey, S. J., Feldman, P. A., Redman, R. O., et al. 2000, \apj, 543, L157
\bibitem[Castor(1975)]{Castor75} Castor, J., McCray, R., \&  Weaver, R. 1975, \apj, 200,  L107
\bibitem[Castets(1982)]{Castets82} Castets, A., Langer, W. D., \&  Wilson, R. W. 1982, \apj, 262,  590
\bibitem[Condon(1992)]{Condon92} Condon, J. J.  ARA\&A, 30, 575
\bibitem[Condon(1998)]{Condon98} Condon, J. J., Cotton, W. D, Greisen, E. W., et al., 1998, AJ, 115, 1693
\bibitem[Chen(2013)]{Chen13} Chen, Y., Zhou, P., \& Chu, You-Hua. et al. 2013, \apj, 769,  L16
\bibitem[Churchwell(2002)]{Churchwell02} Churchwell, E. 2012, ARA\&A, 40, 27
\bibitem[Churchwell(2006)]{Churchwell06} Churchwell, E., Povich, M. S., Allen, D., et al. 2006, \apj, 649, 759
\bibitem[Churchwell(2007)]{Churchwell07}Churchwell, E., Watson, D. F., Povich, M. S., et al. 2007, \apj, 670, 428
\bibitem[Dirienzo(2012)]{Dirienzo12} Dirienzo, W. J., Indebetouw, R., \& Brogan, C. 2012, \apj, 144, 173
\bibitem[Deharveng(2010)]{Deharveng10} Deharveng, L., Schuller, F., Anderson, L. D., et al. 2010, A\&A, 523, 6
\bibitem[Dyson(1980)]{Dyson80} Dyson, J. E., \& Williams, D. A. 1980, Physics of the interstellar medium, ed. Dyson, J. E. \& Williams, D. A.
\bibitem[Egan(1998)]{Egan98} Egan, M. P., Shipman, R. F., Price, S. D., et al.  1998, \apj, 495, L199
\bibitem[Faundez(2004)]{Faundez04} Fa\'{u}ndez, S., Bronfman, L., Garay, G., et al. 2004, A\&A, 426, 97
\bibitem[Fazio(2004)]{Fazio} Fazio, G. G., Hora, J. L., Allen, L. E., et al. 2004, ApJS, 154, 10
\bibitem[Fich(1989)]{Fich89} Fich, M., Blitz, L., \& Stark, A. A., 1989, \apj, 342, 272
\bibitem[Fiege(2000)]{Fiege00} Fiege, J. D., \& Pudritz, R. E., 2000, MNRAS, 311, 85
\bibitem[Foster(2012)]{Foster12} Foster, J. B., Stead, J. J., Benjamin, R. A., Hoare, M. G., \& Jackson, J. M. 2012, \apj, 751, 157
\bibitem[Foster(2014)]{Foster14} Foster, J. B., Arce, H. G., Kassis, M. et al. 2014, \apj, 791, 108
\bibitem[Goodman(2014)]{Goodman14} Goodman, A. A., Alves, J., Beaumont, C. N. et al. 2014,  \apj, 797, 53
\bibitem[Hennebelle(2001)]{Hennebelle01} Hennebelle, P., P\'{e}rault, M., Teyssier, D., \&  Ganesh, S. 2001, A\&A, 365,598
\bibitem[Hunter(1998)]{Hunter98} Hunter, T. R., Neugebauer, G., Benford, D. J. et al. 1998, \apj, 493, 97
\bibitem[Inoue(2001)]{Inoue01} Inoue, A. K. 2001, AJ, 122, 1788
\bibitem[Jackson(2010)]{Jackson} Jackson, J. M., Finn, S. C., Chambers, E. T., et al. 2010, \apj, 719, L185
\bibitem[Kainulainen(2011)]{Kainulainen} Kainulainen, J., Alves, J., Beuther, H., Henning, T., \& Schuller, F. 2011,  A\&A, 536, 48
\bibitem[Kolpak(2003)]{Kolpak} Kolpak, M. A., Jackson, J. M., Bania, T. M., Clemens, D. P., \& Dickey, J. M. 2003, \apj, 582, 756
\bibitem[Kraemer(1999)]{Kraemer} Kraemer, K. E., Deutsch, L. K., Jackson, J. M., et al. 1999, \apj, 516, 817
\bibitem[Kurayama(2011)]{Kurayama} Kurayama, T., Nakagawa, A., Sawada-Satoh, S., et al. 2011, PASJ, 63, 513
\bibitem[Leger(1984)]{Leger1984} Leger, A., \&  Puget, J. L. 1984, A\&A, 137,5
\bibitem[Liu(2014)]{Liu14} Liu, Xiao-Lan, Wang, Jun-Jie, \& Xu, Jin-Long.  2014, MNRAS, 443, 2264
\bibitem[Liu(2013)]{Liu13} Liu, T., Wu, Y, \& Zhang, H., 2013, \apj, 776, 29
\bibitem[McCray(1983)]{McCray83} McCray, R., 1983, Highlights of Astronomy, 6, 565
\bibitem[Miralles(1994)]{Miralles94} Miralles, M. P., Rodriguez, L. F., \& Scalise, E. 1994, \apjs, 92, 173
\bibitem[Mookerjea(2007)]{Mookerjea07} Mookerjea, B., Casper, E., Mundy, L. G., \&  Looney, L. W. 2007, \apj, 659, 447
\bibitem[Molinari(1998)]{Molinari98} Molinari, S., Brand, J., Cesaroni, R., Palla, F., \&  Palumbo, G. G. C. 1998, A\&A, 336, 339
\bibitem[Paron(2009)]{Paron09} Paron, S., Cichowolski, S., \& Ortega, M. E. 2009, A\&A, 506, 789
\bibitem[Peretto(2009)]{Peretto09} Peretto, N., \& Fuller, G. A.  2009, A\&A, 2009, 405
\bibitem[Petriella(2010)]{Petriella10} Petriella, A., Paron, S., \& Giacani, E., 2010, A\&A, 513, A44
\bibitem[Pillai(2006)]{Pillai06} Pillai, T., Wyrowski, F., Carey, S. J., \&  Menten, K. M. 2006, A\&A, 450, 569
\bibitem[Pomar¨¨s(2009)]{Pomar¨¨s09} Pomar¨¨s, M., Zavagno, A., Deharveng, L., et al.  2009, A\&A, 494, 987
\bibitem[Ragan(2011)]{Ragan11} Ragan, S. E., Bergin, E. A., \&  Wilner, D.  2011, \apj, 736, 163
\bibitem[Rathborne(2005)]{Rathborne05} Rathborne, J. M., Jackson, J. M., Chambers, E. T., et al. 2005, \apj, 630, L181
\bibitem[Rathborne(2006)]{Rathborne06} Rathborne, J. M., Jackson, J. M., \& Simon, R. 2006, \apj, 641, 389
\bibitem[Rathborne(2007)]{Rathborne07} Rathborne, J. M., Simon, R., \& Jackson, J. M.  2007, \apj, 662, 1082
\bibitem[Rathborne(2008)]{Rathborne08} Rathborne, J. M., Jackson, J. M., Zhang, Q., \& Simon, R. 2008, \apj, 689,1141
\bibitem[Rathborne(2011)]{Rathborne11} Rathborne, J. M., Garay, G., Jackson, J. M., et al. 2011, \apj, 741, 120
\bibitem[Reid(1985)]{Reid} Reid, M. J., \& Ho, P. T. P. 1985, \apj, 288, 17
\bibitem[Rieke(2004)]{Rieke} Rieke, G. H., Young, E. T., Engelbracht, C. W., et al. 2004, ApJS, 154, 25
\bibitem[Sanhueza(2010)]{Sanhueza10} Sanhueza, P., Garay, G., Bronfman, L., et al. 2010, \apj, 715, 18
\bibitem[Sakai(2013)]{Sakai13} Sakai, T., Sakai, Nami., Foster, J. B., et al. 2013, \apj, 775, L31
\bibitem[Scoville(1986)]{Scoville86} Scoville, N. Z., Sargent, A. I., Sanders, D. B., et al. 1986, \apj, 303, 416
\bibitem[Schuller(2009)]{Schuller09} Schuller, F.; Menten, K. M.; Contreras, Y. et al. 2009, A\&A, 504, 415
\bibitem[Shepherd(2004)]{Shepherd04} Shepherd, D. S., N\"{u}rnberger, D. E. A., \& Bronfman, L. 2004, \apj, 602, 850
\bibitem[Shepherd(2007)]{Shepherd07} Shepherd, D. S., Povich, M. S., Whitney, B. A., et al. 2007, \apj, 669, 464
\bibitem[Simon(2006)]{Simon06} Simon, R., Rathborne, J. M., Shah, R. Y., Jackson, J. M., \&  Chambers, E. T. 2006, \apj, 653, 1325
\bibitem[Simpson(2012)]{Simpson12}  Simpson, R. J., Povich, M. S., Kendrew, S.  et al. 2012, MNRAS, 424, 2442
\bibitem[Siringo(2009)]{Siringo09} Siringo, G., Kreysa, E., Kov\'{a}cs, A., et al. 2009, A\&A, 497, 945
\bibitem[Smith(2002)]{Smith02} Smith, L. J., Norris, R. P. F., \& Crowther, P. A. 2002,  MNRAS, 337, 1309
\bibitem[Stahler(2005)]{Stahler05} Stahler, S. W., Palla, F., \& Palla, F. 2005, The Formation of Stars (Physics Textbook), 1st edn. (Wiley-VCH), 204
\bibitem[Szymczak(2000)]{Szymczak00} Szymczak, M. \& Kus, A. J. 2000, A\&A, 360, 311
\bibitem[Tackenberg(2012)]{Tackenberg12} Tackenberg, J., Beuther, H., Henning, T.,  et al. 2012, A\&A, 540, 113
\bibitem[Wang(2015)]{Wang15} Wang, K., Testi, L., Ginsburg, A., et al. 2015, MNRAS, 450, 4043
\bibitem[Watson(2008)]{Watson08} Watson, C., Povich, M. S., Churchwell, E. B., et al.  2008, \apj, 681, 1341
\bibitem[Wu(2007)]{Wu07} Wu, Y., Henkel, C., Xue, R., Guan, X., \& Miller, M. 2007, \apj, 669, L37
\bibitem[Yonekura(2005)]{Yonekura05} Yonekura, Y., Asayama, S., Kimura, K.,  et al. 2005,  \apj, 634, 476
\end{thebibliography}
\end{document}